\documentclass[10pt,aps,pra,twocolumn,showpacs,superscriptaddress,preprintnumbers,amsmath,nofootinbib]{revtex4-2}

\usepackage[utf8]{inputenc} 
\usepackage{xcolor}
\usepackage{dcolumn}
\usepackage{hyperref}
\usepackage{graphicx}
\usepackage[capitalise]{cleveref} 
\usepackage{acronym}
\usepackage{amssymb}
\usepackage{booktabs}
\usepackage{gensymb}

\makeatletter
\newcommand {\crefext}[2]{\csname cref@#1@format\endcsname{#2}{}{}}
\newcommand {\Crefext}[2]{\csname Cref@#1@format\endcsname{#2}{}{}}
\makeatother

\crefname{pluralequation}{Eqs.}{Eqs.}

\DeclareMathOperator{\diag}{diag}
\DeclareMathOperator{\tr}{Tr}
\DeclareMathOperator{\im}{Im}
\DeclareMathOperator{\re}{Re}

\newcommand{\PreserveBackslash}[1]{\let\temp=\\#1\let\\=\temp}
\newcolumntype{C}[1]{>{\PreserveBackslash\centering}p{#1}}

\newcommand{\kdet}{k_\text{det}}
\newcommand{\hatp}[1]{\widehat{#1}}
\newcommand{\mz}[2]{$(#1,#2)$}
\newcommand{\m}[2]{\scriptsize $(#1,#2)$\vphantom{$\hatp{10}'$}}
\newcommand{\mt}[3]{\scriptsize $(#1,#2)$\vphantom{$\hatp{10}'$}}
\newcommand{\N}{\text{N}}

\allowdisplaybreaks

\begin{document}

\preprint{SISSA 06/2024/FISI}

\title{Finite modular symmetries and the strong CP problem}

\author{J. T. Penedo}
\affiliation{INFN Sezione di Roma Tre, Via della Vasca Navale 84, 00146, Roma, Italy}
\author{S. T. Petcov}
\altaffiliation{Also at: Institute of Nuclear Research and Nuclear Energy,
Bulgarian Academy of Sciences, 1784 Sofia, Bulgaria.}
\affiliation{SISSA/INFN, Via Bonomea 265, 34136 Trieste, Italy}
\affiliation{Kavli IPMU (WPI), UTIAS, The University of Tokyo, Kashiwa, Chiba 277-8583, Japan}

\begin{abstract}
Recently, it was shown that modular symmetry may solve the strong CP problem without axions, by producing a vanishing QCD angle while generating a large quark CP violation phase. We extend this framework to finite modular groups, systematically identifying the allowed mass textures. We find quark fields must furnish 1D representations and scan the minimal model landscape.
\end{abstract}

\maketitle

\section{Introduction}
\label{sec:intro}

In an effort to emulate the undeniable success of the gauge principle in the \ac{SM},
flavour symmetries have been put forward to explain the puzzling observed structures of masses and mixing in the quark and lepton sectors of the theory~\cite{Ishimori:2010au,Altarelli:2010gt,King:2014nza,Feruglio:2015jfa,Tanimoto:2015nfa,Petcov:2017ggy,Xing:2020ijf,Feruglio:2019ybq}. The curious avenue of finite modular flavour symmetries~\cite{Feruglio:2017spp} has attracted attention in the last years, due in part to the promising possibility of shaping flavour patterns without the need to invoke large numbers of exotic scalar fields, with aligned \acp{VEV}.
Modular flavour models allow one to 
determine fermion masses, mixing angles and Dirac and Majorana \ac{CPV} phases in terms of a limited number of parameters;
generate fermion mass hierarchies~\cite{Criado:2019tzk,King:2020qaj,Feruglio:2021dte,Novichkov:2021evw,Novichkov:2022wvg,Kuranaga:2021ujd}; and
provide a unified source of \ac{gCP} and flavour symmetry breaking~\cite{Baur:2019kwi,Novichkov:2019sqv},
to name key features of the framework (see~\cite{Kobayashi:2023zzc,Ding:2023htn} for recent reviews).

More recently, it has been shown~\cite{Feruglio:2023uof} that modular symmetries can play a role in unveiling yet another puzzle of the \ac{SM} --- the strong CP problem --- related to the non-observation of an electric dipole moment for the neutron. Solutions to this conundrum usually fall into two categories: i) those promoting the relevant QCD angle $\bar \theta$ to a dynamical field --- the axion~\cite{Peccei:1977hh,Peccei:1977ur,Wilczek:1977pj,Weinberg:1977ma} (see~\cite{DiLuzio:2020wdo} for a review),%
\footnote{See also~\cite{Kobayashi:2020oji,Higaki:2024jdk} for a possible modular origin of the axion.}
 and ii) those breaking an assumed CP symmetry in a way that produces a large \ac{CPV} phase $\delta$ in the \ac{CKM} quark mixing matrix while simultaneously ensuring $\bar \theta = 0$. In what follows, we will focus on the latter possibility.
Here, 
\begin{equation} \label{eq:thetabar}
\bar\theta = \theta_\text{QCD} + \arg \det M_u M_d\,,
\end{equation}
%
where $\theta_\text{QCD}$ is the coefficient of the topological term in the QCD Lagrangian, while $M_q$ ($q=u,d$) is the $q$-type quark mass matrix. The assumed (g)CP symmetry of the theory can provide $\theta_\text{QCD} = 0$. The requirement on the flavour model is then to generate a viable $\delta$ from $M_u$ and $M_d$ while keeping the product of their determinants real.
This could be achieved in models with vector-like quarks (see e.g.~\cite{Alves:2023ufm}), most famously via Nelson-Barr constructions~\cite{Nelson:1983zb,Barr:1984qx} (see also~\cite{Bento:1991ez,Dine:2015jga,Cherchiglia:2020kut,Valenti:2021rdu}).%
\footnote{For other solutions of the strong CP problem based on the spontaneous breaking of CP symmetry, see~\cite{Hiller:2001qg,Harnik:2004su,Cheung:2007bu,Antusch:2013rla,Vecchi:2014hpa}.}
We instead follow the minimal approach put forward in Ref.~\cite{Feruglio:2023uof}, wherein modular symmetry is elegantly responsible for the necessary flavour textures, and extend this approach to the finite modular groups.

\vskip -8mm
${ }$

\section{Setup}
\label{sec:setup}

Consider the typical $\mathcal{N}=1$ global SUSY modular framework
with a \ac{gCP} symmetry~\cite{Novichkov:2019sqv}  (see e.g.~\cite{Penedo:2018nmg,Novichkov:2018ovf,Novichkov:2020eep} for further details).%
\footnote{A CP-conserving mechanism of SUSY breaking is also assumed, resulting in a real gluino mass so that $\bar\theta =0$ is preserved.}
The theory is taken to be invariant under the whole modular group $SL(2, \mathbb{Z})$. Matter fields $\psi$ instead transform in representations of a \emph{finite} modular group of a certain level $N$ --- which plays the role of a flavour symmetry --- and are assigned modular weights $k_\psi$. The inhomogeneous finite groups $\Gamma_N$ are isomorphic to the well-known permutation groups $S_3$~\cite{Kobayashi:2018vbk,Meloni:2023aru}, $A_4$~\cite{Feruglio:2017spp,Criado:2018thu,Novichkov:2018yse}, $S_4$~\cite{Penedo:2018nmg,Novichkov:2018ovf,Criado:2019tzk} and $A_5$~\cite{Novichkov:2018nkm,Criado:2019tzk}, while the homogeneous finite groups $\Gamma'_N$ are isomorphic to ``double covers'' of these groups~\cite{Liu:2019khw,Novichkov:2020eep,Liu:2020akv,Wang:2020lxk,Yao:2020zml}. One can hence focus on $\Gamma_N' \supseteq \Gamma_N$ without loss of generality.

We are interested in simple scenarios, where Yukawa and mass matrices depend only on modular forms of level $N$ --- known functions of a single complex scalar field $\tau$ (the modulus) --- and on a small number of coupling constants in the holomorphic superpotential $W$. The flavour structure is determined once $\tau$ acquires a \ac{VEV}, which breaks both modular and CP symmetries. 
For simplicity, we consider a minimal-form Kähler potential,
\begin{equation} \label{eq:kahler}
K = - \Lambda_K^2 \log (2\im \tau) + \sum_\psi |\psi|^2 (2 \im \tau)^{-k_\psi}\,,
\end{equation}
%
where $\Lambda_K$ has mass dimension one. While not the most general choice~\cite{Chen:2019ewa,Chen:2021prl}, it does not hinder the envisioned solution to the strong CP problem, as $\bar\theta$ is holomorphic and thus constrained by modular invariance, being insensitive to $K$, as shown in~\cite{Feruglio:2023uof}.
We further take the MSSM Higgs doublets $H_{u,d}$ to transform trivially under the modular group. 

\vskip 2mm

The quark doublets $Q$ and quark singlets $u^c$ and $d^c$ 
furnish 3-dimensional representations of $\Gamma'_N$, which may be reducible in general. Under a modular transformation $\gamma = \left(\begin{smallmatrix} a & b \\ c & d \end{smallmatrix}\right) \in SL(2,\mathbb{Z})$, one has
\begin{equation} \label{eq:modtrans}
\begin{aligned}
Q_i \,&\xrightarrow{\gamma}\, (c\tau + d)^{-k_i} \,\rho_{ij}(\gamma)\, Q_j \,,\\[1mm]
q^c_i \,&\xrightarrow{\gamma}\, (c\tau + d)^{-k^c_i} \,\rho^c_{ij}(\gamma)\, q^c_j \,,
\end{aligned}
\end{equation}
%
for a given sector $q=u,d$.
In the reducible case, the unitary representation matrices $\rho$ and $\rho^c$ are block diagonal.

It can be shown that the determinant of the quark mass matrix $M_q$ is a singlet (i.e.~1D) modular form of weight $\kdet^q$, with $\kdet^q \equiv \sum_i k_i + k^c_i$ (see~\cref{app:det} for explicit derivations).
Immediately, it becomes clear that if $\kdet^q < 0$, then $\det M_q = 0$, as there are no non-zero negative-weight modular forms. In this study, we reject quark mass matrices with zero determinant, as they imply a massless quark.%
\footnote{%
In the presence of a massless quark, an additional mechanism might be invoked to generate the up- or down-quark masses, such as SUSY-breaking effects or higher-dimensional operators, along the lines of what is described in Refs.~\cite{Feruglio:2021dte,Petcov:2023vws}. The latter possibility may require non-trivial weights for $H_{u,d}$. We do not consider such possibilities here.}
The condition $\det M_q \neq 0$ then implies that both $\kdet^u \geq 0$ and $\kdet^d \geq 0$.

To implement the solution to the strong CP problem proposed in Ref.~\cite{Feruglio:2023uof}, one requires the cancellation of the QCD anomaly $A$ of the modular symmetry, which in our case simply reads%
\footnote{%
\cref{eq:A} holds also when the Higgs doublets carry cancelling non-zero modular weights, i.e.~$k_{H_u} = -k_{H_d}$, since $\kdet^q = 3k_{H_q}+ \sum_i k_i + k^c_i$.
We thank M.~Parriciatu for this observation.}
%
\begin{equation} \label{eq:A}
A \,=\, \kdet^u + \kdet^d \stackrel{!}{=} 0\,.
\end{equation}
%
The vanishing of $A$ guarantees that $\det M_u \det M_d$ 
is a modular form of zero weight, hence a constant, independent of the only source of CP breaking, namely the \ac{VEV} of~$\tau$.
Moreover, it is a real constant, given the assumed \ac{gCP} invariance of the theory.
Given the above, the vanishing of $A$ implies the separate vanishing of the modular weights of the determinants of both quark mass matrices,
\begin{equation} \label{eq:zerodets}
\kdet^u \stackrel{!}{=} 0\,,\qquad
\kdet^d \stackrel{!}{=} 0\,.
\end{equation}
%
In the next section, we will look into the consequences of this restriction.

\vskip 2mm

Note that these results already show that
this mechanism of solving the strong CP problem is {\it incompatible} with the mechanism of generating fermion mass hierarchies put forward in Ref.~\cite{Novichkov:2021evw}. This is because the latter is based on suppressing entries of fermion mass matrices via powers of a small $\tau$-dependent quantity, $\epsilon(\tau)$. The determinant of each mass matrix is then inevitably proportional to some power of $\epsilon(\tau)$ and is necessarily $\tau$-dependent, carrying a non-zero modular weight, in contradiction with~\cref{eq:zerodets}.

\section{Allowed textures}
\label{sec:textures}
\subsection{Single sector}
\label{sec:single}

Asking for vanishing determinant weights while allowing for non-zero determinants is a rather restrictive requirement. We analyse this constraint systematically by first focusing on a single quark sector $q$. In~\cref{sec:both}, we discuss the compatibility of different structures for $M_u$ and $M_d$.

We start by associating each entry of the relevant mass matrix $M_q$ in the \ac{LR} convention to a weight $k_{ij}^q \equiv k_i + k^{q^c}_j$, according to
\begin{equation}
M \,:\, \begin{pmatrix}
k_{11} & k_{12} & k_{13} \\
k_{21} & k_{22} & k_{23} \\
k_{31} & k_{32} & k_{33}
\end{pmatrix} \,,
\end{equation}
%
where we have omitted the label $q=u,d$ for readability.
We then have
\begin{equation} \label{eq:kdet}
\kdet \,=\, 
k_{11} + k_{22} + k_{33}\,=\, 
k_{12} + k_{23} + k_{31}\,=\, 
\ldots \,\stackrel{!}{=} \, 0
\,.
\end{equation}
%
Note further that if $k_{ij} < 0$, then the corresponding mass matrix element $M_{ij}$ vanishes, since there are no modular forms of negative weight, and therefore no modular forms available to produce a modular invariant which can contribute to this entry.

The trivial solution to~\cref{eq:kdet} is to have all $k_{ij} = 0$.
Then, given the assumed \ac{gCP} symmetry, broken only by the \ac{VEV} of $\tau$,
CP would be conserved in the quark sector as the elements 
of $M$ would be $\tau$-independent CP-conserving constants.
At the same time, however, we must generate a non-trivial \ac{CPV} \ac{CKM} phase $\delta$. 
Therefore, some $k_{ij}$ must be non-zero. The non-zero ones cannot all be positive, otherwise there is no way of satisfying~\cref{eq:kdet}. Hence, a way to progress methodically is to analyse one by one the cases with a definite number of negative weights $k_{ij}$.

\vskip 5mm
\noindent
{\bf 1 negative $k_{ij}$.}
We are free to permute representations and thus rows and columns such that the negative element is $k_{11}$.%
\footnote{\label{footnote:nD}%
This is true even if we are dealing with representations of dimension higher than 1, in which case some weights are forced to be equal and the action of permutations must simply keep the block structure of the mass matrix.}
Now, since $\kdet = k_{13} + k_{22} + k_{31} = k_{12} + k_{21} + k_{33} \stackrel{!}{=} 0$ and none of these quantities can be negative, one has $k_{22} = k_{33} = 0$. However, since also $k_{11} + k_{22} + k_{33} \stackrel{!}{=} 0$, we find in particular that $k_{11} = 0$, in contradiction with our original assumption.

\vskip 5mm
\noindent
{\bf 2 negative $k_{ij}$.}
If the negative-weight entries do not share a row or column, we can choose $k_{11}, k_{22} < 0$ without loss of generality. Then, one is forced to have $k_{33} > 0$. However, $k_{12} + k_{21} + k_{33} \stackrel{!}{=} 0$ implies negative $k_{12}$ or $k_{21}$, in contradiction with our original assumption.

Instead, one may have the negative-weight entries sharing a row or column. 
Without loss of generality, we can choose $k_{11}, k_{12} < 0$.%
\footnote{If a viable structure is found, then its transpose may also be viable a priori.}
Then, since $\kdet = k_{13} + k_{22} + k_{31} = k_{13} + k_{21} + k_{32}  \stackrel{!}{=} 0$ and none of these quantities can be negative, we have $k_{13} = k_{21} = k_{22} = k_{31} = k_{32} = 0$. Then, $k_{11} + k_{22} + k_{33} \stackrel{!}{=} 0$ implies $k \equiv k_{11}= - k_{33}$, while $k_{12} + k_{21} + k_{33} \stackrel{!}{=} 0$ leads to $k_{12} = - k_{33} = k$. Finally, $k_{11} + k_{23} + k_{32} \stackrel{!}{=} 0$ fixes $k_{23} = - k_{11} = -k$, so that we are left with the viable (and novel) mass matrix weight structure
\begin{equation} \label{eq:v1}
M \,:\, 
\begin{pmatrix}
k & k & 0 \\
0 & 0 & -k \\
0 & 0 & -k
\end{pmatrix} 
 \,,\quad k<0\,,
\end{equation}
%
up to permutations of rows and columns and up to transposition. We can do away with the transposition ambiguity by simply allowing for positive values of $k$.

\Cref{eq:v1} implies the mass matrix has the form
\begin{equation} \label{eq:v1f}
M =  v_q
\begin{pmatrix}
0 & 0 & \alpha_{13} \\[1mm]
\alpha_{21} & \alpha_{22} & \alpha_{23}\,\mathcal{Y}_1^{(|k|)} \\[1mm]
\alpha_{31} & \alpha_{32} & \alpha_{33}\,\mathcal{Y}_2^{(|k|)} 
\end{pmatrix} \,,
\end{equation}
%
also up to permutations and transposition.
Here, $v_q$ is the \ac{VEV} of $H_q$, $\alpha_{ij}$ are real CP-conserving constants (under our working assumption that the only source of \ac{CPV} is the \ac{VEV} of $\tau$), and the $\mathcal{Y}_i$
are obtained from modular forms of weight $|k|$.
The determinant of the mass matrix in~\cref{eq:v1f} is given by
\begin{equation}
\det M = v_q^3\,\alpha_{13}\,(\alpha_{21} \alpha_{32} - \alpha_{22} \alpha_{31})\,.
\end{equation}
%
It is non-zero in general, as well as $\tau$-independent, as anticipated.
The form of $M_q$ in~\cref{eq:v1f} is directly obtained, for example, if
 $Q_i$ and $q^c_i$ are trivial singlets of the finite modular group $\Gamma^\prime_N$, with $\rho_{ij} = \rho^c_{ij} = \delta_{ij}$ in~\cref{eq:modtrans}, and if trivial-singlet modular forms are available at weight $|k|$. In the presence of other representations, additional constraints are imposed on the $\alpha_{ij}$.

\vskip 5mm
\noindent
{\bf 3 negative $k_{ij}$.}
There are four possibilities to consider, up to permutations and transposition. Either i) all negative-weight entries share a row or column, ii) none of these entries share a row or column, iii) exactly one pair of these entries shares a row or column, or iv) exactly two pairs of these entries shares a row or column.

We exclude case i), since in this case $M$ has a vanishing row or column, leading to a massless quark.
In case ii), we can take $k_{11}, k_{22}, k_{33} < 0$ without loss of generality. Immediately, we find $\kdet = k_{11} + k_{22} + k_{33} < 0$, excluding this scenario.
In case iii), we can take $k_{11}, k_{21}, k_{32} < 0$ without loss of generality. Now, since $\kdet = k_{13} + k_{22} + k_{31} \stackrel{!}{=} 0$ and none of these quantities can be negative, one has $k_{13} = 0$. This leads to a contradiction, as it implies $\kdet = k_{13} + k_{21} + k_{32} < 0$, excluding this scenario.

The only viable option is case iv), for which we can take $k_{11}, k_{12}, k_{21} < 0$ without loss of generality. Then, we have $k_{13} + k_{22} + k_{31} = \ldots \stackrel{!}{=} 0$ and the following relations:
\begin{equation}
\begin{aligned}
& k_{13} = k_{22} = k_{31} = 0 \,,\\
& k_{21} = - k_{32} \equiv k \,,\\
& k_{12} = - k_{23} \equiv k' \,,\\
& k_{12} + k_{21} = - k_{33} =  k_{11} = k+k' \,.
\end{aligned}
\end{equation}
%
These relations define,
up to permutations of rows and columns,
a viable family of (weight) structures 
\begin{equation} \label{eq:v2}
M \,:\, 
\begin{pmatrix}
k+k' & k' & 0 \\
k & 0 & -k' \\
0 & -k & -k-k'
\end{pmatrix}
 \,,\quad k,k'<0\,,
\end{equation}
%
corresponding to a mass matrix of the form
\begin{equation} \label{eq:v2f}
M = 
v_q\,\begin{pmatrix}
0 & 0 & \alpha_{13} \\
0 & \alpha_{22} & \alpha_{23}\,\mathcal{Y}_1^{(|k'|)} \\
\alpha_{31} & \alpha_{32}\,\mathcal{Y}_2^{(|k|)} & \alpha_{33}\,\mathcal{Y}_3^{(|k+k'|)}
\end{pmatrix} \,.
\end{equation}
%
Also here, the $\alpha_{ij}$ are real CP-conserving constants, while the $\mathcal{Y}_i^{(w)}$ are necessarily
obtained from singlet (1D) modular forms of weight $w$ (see also~\cref{sec:compat}).
In this case, one has
\begin{equation}
\det M = - v_q^3\,\alpha_{13}  \alpha_{22} \alpha_{31}\,,
\end{equation}
%
which is $\tau$-independent and does not vanish in general.
As before, if both quark fields and the available modular forms furnish trivial singlets of $\Gamma'_N$, the form in~\cref{eq:v2f} directly follows.  Instead, in the presence of other representations or in the absence of modular forms of the requisite weight, one or several of the $\alpha_{ij}$ may vanish.

\vskip 2mm

Note that, given the permutation freedom, we can relax the assumption 
of negative $k,k'$ and take $k,k' \in \mathbb{R}$, with the caveat that 
at least one of them must be non-zero. If exactly one of them vanishes 
or if $k=-k'$, one can check that this structure reduces to that of~\cref{eq:v1}.
The textures considered in Ref.~\cite{Feruglio:2023uof} are contained in the family of structures described by~\cref{eq:v2,eq:v2f} with non-zero $k$, $k'$ and $k+k'$, while the example explored numerically therein corresponds to a subcase with $k = k'$.

\vskip 5mm
\noindent
{\bf 4 negative $k_{ij}$.}
Since in this case some pair of negative-weight entries must share either a row or column, we take $k_{11}, k_{12}<0$ without loss of generality. Then, $k_{13}\geq 0$ to avoid a massless quark. A third negative-weight entry may either i) share a column with $k_{11}$ or $k_{12}$, say $k_{21} <0$  without loss of generality, or ii) be in the third column.

In case i), $k_{22}, k_{31} \geq 0$ to avoid a massless quark, and in fact, due to $k_{13} + k_{22} + k_{31} \stackrel{!}{=} 0$, we have $k_{13} = k_{22} = k_{31} = 0$. Then, no matter where we choose to place the fourth negative-weight entry ($k_{23}$, $k_{33}$ or $k_{32}$), we will run into a conflict with the condition of vanishing determinant, since either $k_{12} + k_{23} + k_{31}$, $k_{12} + k_{21} + k_{33}$ or $k_{13} + k_{21} + k_{32}$ will be negative.

Case ii) is similarly excluded. It corresponds to $k_{23}, k_{33} < 0$, since any other choice is covered by case i). Again, due to $k_{13} + k_{22} + k_{31} \stackrel{!}{=} 0$, we have $k_{13} = k_{22} = k_{31} = 0$. This immediately conflicts with the condition $k_{11} + k_{22} + k_{33} \stackrel{!}{=} 0$ of vanishing determinant.

\vskip 5mm
\noindent
{\bf 5 (or more) negative $k_{ij}$.}
In this case, the reasoning follows that of case i) for 4 negative $k_{ij}$. Without loss of generality, one has $k_{11}, k_{12}, k_{21} < 0$ and $k_{13} = k_{22} = k_{31} =0$. For the same reasons as above, it is not possible to place the fourth negative-weight entry without running into a contradiction. {\it This reasoning excludes any higher number of negative-weight entries.}

\vskip 5mm
\noindent
{\bf 6 negative $k_{ij}$.}
An alternative way to discard this scenario follows. To avoid a massless quark, the entries with non-negative weight cannot share any row or column. Then, one can take the off-diagonal entries to be the negative-weight ones without loss of generality. Then $k_{12} + k_{23} + k_{31} < 0$ conflicts with the condition of vanishing determinant weight.

\vskip 5mm
\noindent
{\bf $7$ or more negative $k_{ij}$.}
An alternative way to discard these cases: $M$ has too many vanishing entries and a massless quark is unavoidable.

\subsection{Both sectors}
\label{sec:both}

We have seen that, for three families, the viable weight textures in a certain quark sector (up or down) are summarized by~\cref{eq:v2}, but with real $k,k'$ and $(k,k') \neq (0,0)$. 
Let us now discuss the compatibility between possible $M_u$ and $M_d$ textures. The two quark sectors are connected by the doublets  $Q_i$ of field weight $k_i$. One has $k_{ij}^q = k_i + k^{q^c}_j$ ($q = u,d$), which implies that
\begin{equation} \label{eq:diffs}
k_{ik}^u - k_{jk}^u = k_{ik}^d - k_{jk}^d = k_i - k_j\,,
\end{equation}
%
i.e.~the differences $k_{ik}^q - k_{jk}^q$ are $q$-independent. Therefore, suppose that we fix
\begin{equation} \label{eq:Mu}
M_u \,:\, \begin{pmatrix}
k_u+k_u' & k_u' & 0 \\
k_u & 0 & -k_u' \\
0 & -k_u & -k_u-k_u'
\end{pmatrix} \,,
\end{equation}
%
in the \ac{LR} convention, using up our permutation freedom of the $Q_i$ and $u^c_i$.
The only remaining freedom is in permuting the columns of $M_d$, which will however not affect quark masses or standard (left-handed) quark mixing. Then, using the $q$-independence of the differences in~\cref{eq:diffs}, one constrains the weight structure in the down-quark sector to
\begin{equation} \label{eq:Md}
M_d \,:\,
\begin{pmatrix}
a+k_u' & b+k_u' & c+k_u' \\
a & b & c \\
a-k_u & b-k_u & c-k_u
\end{pmatrix}
\,,
\end{equation}
%
with $a,b,c$ as yet undetermined. This structure  must match some permutation of the most general viable one in~\cref{eq:v2}. This means that each of the rows in~\cref{eq:Md} will contain at least one vanishing entry. We can set $c+k_u' = 0$, $b =0$ and $a - k_u = 0$ without loss of generality, since we are still free to permute the columns. We thus find that $M_d$ necessarily has the same weight structure as $M_u$ in~\cref{eq:Mu}, i.e.
\begin{equation} \label{eq:Mud}
\begin{aligned}
M_u \,&:\, \begin{pmatrix}
k+k' & k' & 0 \\
k & 0 & -k' \\
0 & -k & -k-k'
\end{pmatrix} \,, \\[2mm]
M_d \,&:\,
\begin{pmatrix}
k+k' & k' & 0 \\
k & 0 & -k' \\
0 & -k & -k-k'
\end{pmatrix}
\,,
\end{aligned}
\end{equation}
%
with shared $k,k'$ (real, at least one non-zero).

The structures of $M_u$ and $M_d$ in~\cref{eq:Mud}
correspond to the most general result.
They are valid up to weak basis changes, i.e.~up to sector-independent permutations of columns and up 
to {\it simultaneous} permutations of rows in both sectors.
Importantly,  all such permutations are physically inconsequential.
Indeed, in the \ac{LR} convention we use, 
we have the singular value decomposition
$M_q = V_L^q D_q V_R^{q\dagger}$, 
where $V_{L,R}^q$ are unitary matrices and  $D_q$
are diagonal matrices of quark masses.  
Note that the unitary matrices $V_L^q$ diagonalize
$M_q M_q^\dagger$, whose eigenvalues are the squares of the quark masses.
The quark mixing matrix is given by $U_{\rm CKM} = V_L^{u\dagger} V^d_L$.
The permutation of columns $i$ and $j$ of $M_q$ can be realized by a real 
orthogonal matrix $P_{ij}$ acting on $M_q$ on the right: $M_q\to M_q P_{ij}$. 
However,  the products $M_qM^\dagger_q$ determining
$V_L^q$ (and thus $U_{\rm CKM}$) clearly do not depend on $P_{ij}$.
The simultaneous permutation of the  $i$ and $j$ rows of $M_u$ and $M_d$ 
is instead realized by the same matrix $P_{ij}$ acting on both  $M_u$ and $M_d$ on the left:
$M_q \to P_{ij}M_q$.
However, $M_{q}M^\dagger_{q}$ 
and $P_{ij}M_{q}M^\dagger_{q}P^T_{ij}$ have the same eigenvalues,
and both $V^{u}_L$ and $V^{d}_L$ are modified by the additional factor $P_{ij}$
on the left, $V^{q}_L \to P_{ij}V^{q}_L$, which cancels 
in the expression for $U_{\rm CKM}$.

\subsection{Compatibility with field representations}
\label{sec:compat}
The structures summarized in~\cref{eq:Mud} can of course be realized by appropriately 
choosing the weights of quark fields furnishing only 1D representations 
of $\Gamma'_N$, i.e.~$Q,u^c, d^c \sim \mathbf{1}^* \oplus \mathbf{1}^* \oplus \mathbf{1}^*$,%
\footnote{%
Here and in what follows, $\mathbf{r}^*$ denotes some unspecified $\Gamma^{(\prime)}_N$ irrep
of a given dimension. In the case of 
$\Gamma_4 \simeq S_4$, for example, 
$\mathbf{1}^* \in \{\mathbf{1},\mathbf{1}'\}$,
$\mathbf{2}^* = \mathbf{2}$ (the only possibility), and $\mathbf{3}^* \in \{\mathbf{3},\mathbf{3}'\}$.
}
with 
all irreducible representations (irreps) a priori unrelated, apart from the requirement of having enough 
invariants to populate the mass matrix and avoid massless quarks (see also~\cref{sec:1+1+1}).
The presence of irreducible representations of dimension higher than 1 
translates into equating the weights $k_{ij}$ within some blocks of the mass 
matrices (see also~\cref{footnote:nD}).
For instance, if 
$Q \sim \mathbf{3}^*$, then 
$k^q_{1j} \stackrel{!}{=} k^q_{2j} \stackrel{!}{=} k^q_{3j}$ ($j=1,2,3$). 
At once we see that this is incompatible with the general result 
in~\cref{eq:Mud}, since it would require $k=k'=0$. The same reasoning excludes 
any isosinglet triplet representations $u^c,d^c \sim \mathbf{3}^*$.
Therefore, avoiding the all-singlet case, one concludes that a doublet irrep
must be present.
The associated constraint of equal weights will imply that one of the following happens, independently of whether $Q$, $u^c$ or  $d^c$ furnishes 
$\mathbf{2}^* \oplus \mathbf{1}^*$: either $k=0$, $k' = 0$ or $k+k'=0$. 
In any case, we will be in the situation of~\cref{eq:v1}. 
More explicitly, one has
\begin{equation} \label{eq:Mud2}
M_u \,:\,
\left(\begin{array}{cc|c}
0 & 0 & k \\
0 & 0 & k \\ \hline 
-k & -k & 0 
\end{array}\right) 
\,, \quad
M_d \,:\,
\left(\begin{array}{cc|c}
0 & 0 & k \\
0 & 0 & k \\ \hline 
-k & -k & 0 
\end{array}\right) 
\,,
\end{equation}
%
with real non-zero $k$, up to (physically irrelevant) 
sector-independent permutations of column blocks and simultaneous 
permutations of row blocks. Note that one cannot rely on non-trivial modular 
forms to populate the zero-weight $2\times 2$ block. It follows that if there is a flavour 
doublet among the isodoublet fields, one must have the conjugate doublet 
in the isosinglet fields (and vice versa) as otherwise there is no way of 
producing an invariant in that block, leading to a massless quark. Therefore, 
the only potentially viable case not involving only 1D irreps
is $Q,u^c, d^c \sim \mathbf{2}^*  \oplus \mathbf{1}^*$, with the weight 
structure given in~\cref{eq:Mud2}. We assess 
its viability in the following section.
Potentially viable cases involving only 1D irreps, taking~\cref{eq:Mud} as a starting point,
are analyzed in~\cref{sec:1+1+1}.

\section{The \texorpdfstring{$\mathbf{2}^* \oplus \mathbf{1}^*$}{2+1} case}
\label{sec:2+1}

As explained in~\cref{sec:compat}, the only potentially viable scenario within this setup
that does not rely on using only 1-dimensional irreps corresponds to having all 
three $Q,u^c, d^c \sim \mathbf{2}^*  \oplus \mathbf{1}^*$. Accordingly, 
we denote $Q = Q_2 \oplus Q_1$, $u^c = u^c_2 \oplus u^c_1$ and 
$d^c = d^c_2 \oplus d^c_1$. Then, to avoid a massless quark, we require 
$\mathbf{r}_{Q_2} \otimes \mathbf{r}_{u^c_2} \supset \mathbf{1}$ and 
$\mathbf{r}_{Q_1} \otimes \mathbf{r}_{u^c_1} = \mathbf{1}$. In other words, 
$\mathbf{r}_{u^c_2} = \overline{\mathbf{r}_{Q_2}}$ and 
$\mathbf{r}_{u^c_1} = \overline{\mathbf{r}_{Q_1}}$, where the bar denotes the 
conjugate representation. The analogous consideration applies to the 
down sector.
In short, we are constrained to the assignments (cf.~\cref{eq:Mud2}):
\begin{equation}
\begin{aligned}
Q &\sim (\mathbf{2}_{Q},k_2)\oplus  (\mathbf{1}_{Q},k_1)\,,\\[1mm]
u^c &\sim (\overline{\mathbf{2}_{Q}},-k_2)\oplus  (\overline{\mathbf{1}_{Q}},-k_1)\,,\\[1mm]
d^c &\sim (\overline{\mathbf{2}_{Q}},-k_2)\oplus  (\overline{\mathbf{1}_{Q}},-k_1)\,,
\end{aligned}
\end{equation}
%
where we have defined
$\mathbf{2}_{Q} \equiv \mathbf{r}_{Q_2}$, $\mathbf{1}_{Q} \equiv \mathbf{r}_{Q_1}$.
Moreover, we have $k = k_2 - k_1$ and must have $k_1 \neq k_2$, 
since $k\neq 0$ is needed to generate a non-trivial \ac{CKM} phase.
Finally, only two entries of the mass matrix --- those corresponding to 
weight $|k|$ --- will be populated by modular forms. 
These forms are necessarily (one or more) doublets furnishing a shared 
representation, $Y \sim (\mathbf{2}_Y, |k|)$. One has
\begin{equation} \label{eq:2Y}
\mathbf{2}_Y \,=\,  \begin{cases}
\,\overline{\mathbf{2}_Q} \otimes \mathbf{1}_Q\,,& \text{if } k>0\,,\\[1mm]
\,\mathbf{2}_Q \otimes \overline{\mathbf{1}_Q}\,,& \text{if } k<0\,.
\end{cases}
\end{equation}
%
In~\cref{tab:2forms} we collect the available doublet modular form irreps for each 
of the finite modular groups $\Gamma'_N$ ($N \leq 5$), up to weight $w = 11$. Beyond this weight, one is sure to have 
at least two independent modular forms $Y_{\mathbf{r},j}^{(w)}$ ($j=1,2,\ldots$) furnishing each possible doublet irrep $\mathbf{r}$.

\begin{table}[t]
  \centering
  \begin{tabular}{lcccc}
    \toprule
    Weight $w$ & \quad\,\,$S_3$\,\,\quad{} & \qquad$A_4'$\qquad{} & \quad\,\,$S_4'$\,\,\quad{} & \quad$A_5'$\quad{} \\
    \midrule
  1 & $-$ & $\mathbf{\hat{2}}$ & $-$ & $-$ \\
  2 & $\mathbf{2}$ & $-$ & $\mathbf{2}$ & $-$ \\
  3 & $-$ & $\mathbf{\hat{2}}$, $\mathbf{\hat{2}''}$ & $-$ & $-$ \\
  4 & $\mathbf{2}$ & $-$ & $\mathbf{2}$ & $-$ \\
  5 & $-$ & $\mathbf{\hat{2}}$, $\mathbf{\hat{2}'}$, $\mathbf{\hat{2}''}$ & $\mathbf{\hat{2}}$ & $\mathbf{\hat{2}}$, $\mathbf{\hat{2}'}$ \\
  6 & $\mathbf{2}$ & $-$ & $\mathbf{2}$ & $-$ \\
  7 & $-$ & $\mathbf{\hat{2}}$, $\mathbf{\hat{2}}$, $\mathbf{\hat{2}'}$, $\mathbf{\hat{2}''}$ & $\mathbf{\hat{2}}$ & $\mathbf{\hat{2}}$, $\mathbf{\hat{2}'}$ \\
  8 & $\mathbf{2}$, $\mathbf{2}$  & $-$ & $\mathbf{2}$, $\mathbf{2}$ & $-$ \\
  9 & $-$ & $\mathbf{\hat{2}}$, $\mathbf{\hat{2}}$, $\mathbf{\hat{2}'}$, $\mathbf{\hat{2}''}$, $\mathbf{\hat{2}''}$ & $\mathbf{\hat{2}}$ & $\mathbf{\hat{2}}$, $\mathbf{\hat{2}'}$ \\
  10 & $\mathbf{2}$, $\mathbf{2}$   & $-$ &  $\mathbf{2}$, $\mathbf{2}$ & $-$ \\ 
  11 & $-$ & $\mathbf{\hat{2}}$, $\mathbf{\hat{2}}$, $\mathbf{\hat{2}'}$, $\mathbf{\hat{2}'}$, $\mathbf{\hat{2}''}$, $\mathbf{\hat{2}''}$ & $\mathbf{\hat{2}}$, $\mathbf{\hat{2}}$ & $\mathbf{\hat{2}}$, $\mathbf{\hat{2}}$, $\mathbf{\hat{2}'}$, $\mathbf{\hat{2}'}$ \\
    \bottomrule
  \end{tabular}
  \caption{Doublet modular form irreps available  at a given weight $w\leq 11$ for the $\Gamma_N'$ groups ($N\leq 5$). Hatted (unhatted) irreps are in correspondence with odd (even) weights.}
  \label{tab:2forms}
\end{table}
%

Going over each of the finite modular groups $\Gamma^{(\prime)}_2 \simeq S_3$, $\Gamma'_3 \simeq A_4$, $\Gamma'_4 \simeq S_4$, and $\Gamma'_5 \simeq A_5$, one finds 
only two inequivalent families of textures are allowed.
Namely, 
for a given non-zero integer $k$ and doublet irrep 
$\mathbf{2}_Y$ available at weight $|k|$ (see~\cref{tab:2forms}),
the viable mass matrices read ($q=u,d$)
\begin{equation} \label{eq:2+1}
M_q = \begin{pmatrix}
\alpha_1^q & 0 & 0 \\
0 & \alpha_1^q & 0 \\
0 & 0 & \alpha_2^q
\end{pmatrix}
+ \alpha_3^q 
\begin{cases}
 \begin{pmatrix}
0& 0 & \mathcal{Y}_1^{q} \\
0 &0 & \mathcal{Y}_2^{q} \\
0 & 0 & 0
\end{pmatrix}\,, &\text{if }k>0\,, \\[8mm]
 \begin{pmatrix}
0& 0 & 0 \\
0 &0 & 0 \\
\mathcal{Y}_1^{q} & \mathcal{Y}_2^{q} & 0
\end{pmatrix}\,, & \text{if }k<0 \,,
\end{cases}
\end{equation}
%
in an appropriate weak basis, with
\begin{equation} \label{eq:sumY}
 \mathcal{Y}_i^{q} \equiv \sum_{j=1} g_j^q \left( Y_{\mathbf{2}_Y,j}^{(|k|)} \right)_i
 \quad \text{and }\quad
g_j^q \equiv \frac{\alpha^q_{j+2}}{\alpha_3^q}
\,.
\end{equation}
%
Here, the $\alpha_i^q$ are \emph{real} superpotential parameters, given the imposed gCP symmetry, while the sum in $j$ goes over all available independent modular forms of the selected weight and representation.
Note the correspondence with the structure in~\cref{eq:v1f}, following a row permutation (and transposition, if $k<0$). As anticipated, the presence of doublet representations leads to constraints on the $\alpha_{ij}$, which are dictated by tensor products. The form in~\cref{eq:2+1} may not follow immediately from the Clebsch-Gordan coefficients used in the literature, but is always reachable via field redefinitions (weak basis transformations).

Finally,~\cref{eq:2+1} must be modified to include the effects of canonical normalization.
Given the assumed minimal-form Kähler of~\cref{eq:kahler}, 
bringing the kinetic terms to a standard form leads to a renormalization 
of the quark fields. As a result, the couplings
multiplying the two components of the doublet modular forms 
are modified as follows:
\begin{equation} \label{eq:ta3}
\alpha^q_i \,\to\, \tilde{\alpha}^q_i = 
\alpha^q_i\, ({2\im\tau})^{|k|/2} \quad (i \geq 3)\,.
\end{equation}
%
Correspondingly, one has $\alpha_3^q \to \tilde{\alpha}_3^q$ in~\cref{eq:2+1}, 
while keeping the ratios $g_j^q$ and thus the $\mathcal{Y}_i^q$ unchanged.

\vskip 2mm

The found $\mathbf{2}^*  \oplus \mathbf{1}^*$ textures have a chance of being highly predictive. Aside from the \ac{VEV} of the modulus, 
the mass matrix $M_q$ in each of the two sectors 
depends only on three real constants, in the minimal case where a single modular form is
present in the sum of~\cref{eq:sumY}.
Nevertheless, one must be able to obtain a viable quark \ac{CPV} phase $\delta$.
Recall that values of the modulus
lying on the border of the fundamental domain $\mathcal{D}$ of the 
modular group and on the line $\re\tau = 0$ 
conserve CP~\cite{Novichkov:2019sqv}, while 
all other values of $\tau$ in $\mathcal{D}$ violate CP.
The latter lead to complex, CP-violating
modular forms in the 
quark mass matrices.
Writing the components of the $\mathcal{Y}^{q}$ as
\begin{equation} \label{eq:Yj}
\mathcal{Y}_i^{q} = |\mathcal{Y}_i^{q}|\,e^{i\phi^q_i}\quad (i=1,2)\,,
\end{equation}
%
one may cast each mass matrix in the form
\begin{equation}
M_q = P_q \tilde{M}_q P_q^*\,,\quad
P_q = \diag\left(e^{\pm i\phi_1^q},\,e^{\pm i\phi_2^q},\,1\right)\,,
\end{equation}
%
where $\tilde{M}_q$ is a real matrix and signs agree with the sign of $k$.
The unitary matrix diagonalizing $M_qM^\dagger_q$ 
is then given by $V^q_L = P_q\, O^q_L$, 
where $O^q_L$  is the orthogonal matrix diagonalizing
$\tilde{M}_q \tilde{M}_q^\dagger$.
For the \ac{CKM} matrix we obtain
\begin{equation} \label{eq:UCKM}
U_{\rm CKM} = V_L^{u\dagger} V^d_L = O^{uT}_L P_u^* P_d \,O^d_L\,,
\end{equation}
%
with
\begin{equation}
 P_u^*P_d = \diag\left(e^{\pm i\left(\phi_1^d - \phi_1^u\right)},\,
e^{\pm i\left(\phi_2^d-\phi_2^u\right)},\,1\right)\,.
\end{equation}
%
Thus, a necessary condition for having 
a complex, CP-violating \ac{CKM} matrix reads
\begin{equation} \label{eq:CPVcond}
\phi_1^d - \phi_1^u \neq 0\quad\text{or}\quad
\phi_2^d - \phi_2^u \neq 0\,.
\end{equation}
%
This condition is not fulfilled if a single modular form 
contributes to the mass matrix, as it must be shared by both the up and down sectors, implying $\phi_i^d = \phi_i^u$.
This problem is generically bypassed if more than one doublet form contributes to the $\mathcal{Y}_i^q$.
Unfortunately, as shown in the remainder of this section, even in this case these textures turn out to be too restrictive and cannot lead to phenomenologically viable quark mass matrices.

\subsection{Texture limitations}
\label{sec:limits}

One can show that the potentially viable $\mathbf{2}^* \oplus \mathbf{1}^*$ textures, summarized in~\cref{eq:2+1}, cannot fit quark data. To start, notice that quark mass ratios severely constrain the $\alpha_i^q$. Namely, given the hierarchical structure of quark masses, one finds, independently of the sign of $k$,
\begin{align} \label{eq:dettr1}
&\det M_q= v^3_q\,(\alpha_1^q)^2\,\alpha_2^q  = m^q_1\,m^q_2\,m^q_3 \,,\\[2mm]
&\tr \left(M_qM^\dagger_q\right) = v^2_q\left [2(\alpha_1^q)^2 + (\alpha_2^q)^2 + \beta_q^2 \right ]
\nonumber \\
\label{eq:dettr2}
&\qquad\qquad\quad\,= 
 (m^q_1)^2 + (m^q_2)^2 + (m^q_3)^2 \simeq (m^q_3)^2
\,, \\[2mm]
&\frac{1}{2}\left[\left(\tr\left(M_qM^\dagger_q\right)\right)^2-\tr\left(\left(M_qM^\dagger_q\right)^2\right)\right] 
\nonumber \\
&\qquad= 
v^4_q\,(\alpha_1^q)^2 \left [(\alpha_1^q)^2 + 2(\alpha_2^q)^2 + \beta_q^2 \right ]
\nonumber \\
&\qquad= 
(m^q_1)^2 (m^q_2)^2 + (m^q_1)^2 (m^q_3)^2 + (m^q_2)^2 (m^q_3)^2
\nonumber \\
\label{eq:dettr3}
&\qquad \simeq  (m^q_2)^2 (m^q_3)^2
\,,
\end{align}
%
with $\beta_q \equiv \tilde{\alpha}_3^q \sqrt{\left|\mathcal{Y}_1^q\right|^2  
+ \left|\mathcal{Y}_2^q\right|^2}$.
Here, $m_i^q$ are the $q$-type quark masses, with $m_3^q \gg m_2^q \gg m_1^q$.
Omitting the index $q$ for readability,~\cref{eq:dettr1,eq:dettr2,eq:dettr3} imply
\begin{equation}
\begin{aligned}
m_3 &\simeq v \sqrt{2\alpha_1^2+ \alpha_2^2 + \beta^2}\,, \\
m_2 &\simeq v\left|\alpha_1\right| \sqrt{\frac{\alpha_1^2+ 2\alpha_2^2 + \beta^2}{2\alpha_1^2+ \alpha_2^2 + \beta^2}} \equiv v\left|\alpha_1\right|f(\alpha_i,\beta) \,, \\
m_1 &\simeq v \, \frac{\left|\alpha_1 \alpha_2\right|}{\sqrt{\alpha_1^2+ 2\alpha_2^2 + \beta^2}} \,.
\end{aligned}
\end{equation}
%
Since the function $f(\alpha_i,\beta)$ is bounded between $1/\sqrt{2}$ and $\sqrt{2}$ for any values of the parameters, one has $m_2 \sim v \left|\alpha_1\right|$ up to an $\mathcal{O}(1)$ factor. Hence, given the strongly hierarchical pattern of quark masses, $m_2^2/m_3^2 \ll 1$ implies $\alpha_1^2 \ll \alpha_2^2 + \beta^2$, while
\begin{equation}
\frac{m_1^2}{m_2^2} \sim \frac{\alpha_2^2}{2\alpha_2^2 + \beta^2}
\ll 1 
\end{equation}
%
implies $\alpha_2^2 \ll \beta^2$. This in turn implies $f(\alpha_i,\beta) \simeq 1$ and $m_2 \simeq v \left|\alpha_1\right|$.
Then, $m_3 \simeq v \left|\beta\right|$ and finally
\begin{equation}
m_1 \simeq v \left|\frac{\alpha_1 \alpha_2}{\beta}\right| \simeq v \left|\alpha_2\right| \frac{m_2}{m_3} \,,
\end{equation}
%
to a good approximation. Using the best-fit values of quark mass ratios summarized in~\cite{deMedeirosVarzielas:2023crv}, obtained at an energy scale $M_\text{GUT} = 2 \times 10^{16}$ GeV for $v_u/v_d = 5$ (see also~\cite{Antusch:2013jca,Bjorkeroth:2015ora,Okada:2020rjb}), one can estimate the ratios of the relevant parameters. For the down sector, one finds
\begin{equation}
\begin{aligned}
\left|\frac{\alpha_1^d}{\beta_d}\right| &\simeq \frac{m_s}{m_b} \simeq 1.37\times 10^{-2} \,,\\
\left|\frac{\alpha_2^d}{\beta_d}\right| &\simeq \frac{m_d}{m_s} \simeq 5.05\times 10^{-2} \,,
\end{aligned}
\end{equation}
%
while for the up sector one has
\begin{equation}
\begin{aligned}
\left|\frac{\alpha_1^u}{\beta_u}\right| &\simeq \frac{m_c}{m_t} \simeq 2.69\times 10^{-3} \,,\\
\left|\frac{\alpha_2^u}{\beta_u}\right| &\simeq \frac{m_u}{m_c} \simeq 2.04\times 10^{-3} \,.
\end{aligned}
\end{equation}
%

Let us now focus on the case $k>0$.
By flipping the signs of quark fields, one can parameterize the mass matrices as
\begin{equation} \label{eq:thparam}
M_q \propto
 \begin{pmatrix}
|\alpha_1^q/\beta_q| & 0 &\cos\theta_q \,e^{i \phi_1^q} \\[1mm]
0 &|\alpha_1^q/\beta_q|  &\sin\theta_q \,e^{i \phi_2^q} \\[1mm]
0 & 0 & |\alpha_2^q/\beta_q|
\end{pmatrix}\,,
\end{equation}
%
up to an overall normalization which does not affect quark mixing. 
Even though only two of the four $\phi_i^q$ phases are physical (see~\cref{eq:CPVcond}), we keep all of them for simplicity.
The columns of the matrices $V_L^{(q)}$ determine the elements of the \ac{CKM} matrix, cf.~\cref{eq:UCKM}. For $k>0$, one has
\begin{equation}
\begin{aligned}
\left(V_L\right)_{i1} &\simeq 
\left( \left|\frac{\alpha_2}{\beta}\right| \cos \theta\, e^{i\phi_1} ,\, \left|\frac{\alpha_2}{\beta}\right| \sin \theta\, e^{i\phi_2} ,\, -1
\right)_i \,, \\[2mm]
\left(V_L\right)_{i2} &= 
\left(\sin \theta\, e^{i\left(\phi_1-\phi_2\right)},\, -\cos\theta,\, 0
\right)_i \,.
\end{aligned}
\end{equation}
%
The first column is obtained at leading order in the small expansion parameters $|\alpha_i/\beta|$, while the result for the second column is exact. It follows that
\begin{equation}
\begin{aligned}
\left|V_{us}\right| &= \left| \left(V_L^u\right)_{i1}^*\left(V_L^d\right)_{i2} \right| \\
&\simeq  \left|\frac{\alpha_2^u}{\beta_u}\right| 
  \left|\cos\theta_u\sin\theta_d \,e^{i \varphi}-\cos\theta_d\sin\theta_u\right| 
\,,
\end{aligned}
\end{equation}
%
with $\varphi \equiv \phi_1^d-\phi_1^u-\phi_2^d+\phi_2^u$. The second factor varies in the interval $[0,1]$, implying the upper bound $\left|V_{us}\right| \lesssim m_u / m_c \simeq 0.002$. This is two orders of magnitude smaller than the value $\left|V_{us}\right| \simeq 0.225$~\cite{deMedeirosVarzielas:2023crv} required by quark data.
By a similar procedure, one can also show that
\begin{equation}
\left|\left|V_{ub}\right| - \frac{m_d}{m_s} \right| \lesssim \frac{m_u}{m_c}\,,
\end{equation}
%
i.e.~that $\left|V_{ub}\right|  \in [0.048,\,0.053]$, again in contradiction with the data, which requires the much smaller $\left|V_{ub}\right| \simeq 0.003$.

\vskip 2mm

Finally, the case $k<0$ is also excluded. For negative $k$, one must replace $M_q \to M_q^T$ in~\cref{eq:thparam},
leading to
\begin{equation}
\begin{aligned}
\left(V_L\right)_{i2} &=
\left(-\sin \theta\, e^{i\left(\phi_2-\phi_1\right)},\, \cos\theta,\, 0
\right)_i \,, \\[2mm]
\left(V_L\right)_{i3} &\simeq 
\left( \left|\frac{\alpha_1}{\beta}\right| \cos \theta\, e^{-i\phi_1} ,\, \left|\frac{\alpha_1}{\beta}\right| \sin \theta\, e^{-i\phi_2} ,\, 1
\right)_i \,.
\end{aligned}
\end{equation}
%
This implies that
\begin{equation}
\begin{aligned}
\left|V_{cb}\right| &= \left| \left(V_L^u\right)_{i2}^*\left(V_L^d\right)_{i3} \right| \\
&\simeq  \left|\frac{\alpha_1^d}{\beta_d}\right| 
  \left|\cos\theta_u\sin\theta_d \,e^{i \varphi}-\cos\theta_d\sin\theta_u\right| 
\,,
\end{aligned}
\end{equation}
%
once again leading to an upper bound on the magnitude of a mixing matrix element. In this case one has $\left|V_{cb}\right| \lesssim m_s/m_b \simeq 0.014$, while data requires $\left|V_{cb}\right| \simeq 0.036$, a value more than twice as large.

\vskip 2mm

The above no-go analytical results are corroborated by numerical scans of the $\mathbf{2}^*  \oplus \mathbf{1}^*$ parameter space, treating the $\mathcal{Y}_i^q$ as free quantities. Indeed, the details of the modular forms play essentially no role in the exclusion of this case, as evidenced by the general parameterization of~\cref{eq:thparam}.
These results may then hold true even if we one is dealing with vector-valued modular forms~\cite{Liu:2021gwa}. 
For instance, the modular binary dihedral group $2D_3$, recently studied in 
Ref.~\cite{Arriaga-Osante:2023wnu}, admits only singlet and doublet irreps (four and two of each, respectively).
The corresponding tensor products are such that there always exists a weak basis in which mass matrices take the form of~\cref{eq:2+1}. A $2D_3$ generalization of the $\mathbf{2}^*  \oplus \mathbf{1}^*$ case is then prevented by the analyzed restrictive texture.

\section{The \texorpdfstring{$\mathbf{1}^* \oplus \mathbf{1}^* \oplus \mathbf{1}^*$}{1+1+1} case}
\label{sec:1+1+1}

As shown in the previous sections, only scenarios where
quarks furnish 1D irreps of the finite modular group
have a chance to be phenomenologically viable.
We start from the general weight structure of~\cref{eq:Mud}. Without loss of generality, it can be made to read
\begin{align}  \label{eq:tex1}
M_q \,&:\,
 \begin{pmatrix}
0 & 0 & k \\
0 & 0 & k \\
-k & -k & 0 
\end{pmatrix}\,,\quad k>0\,, \\[2mm]
 \label{eq:tex2}
M_q \,&:\,
 \begin{pmatrix}
0 & k' & k' \\
-k' & 0 & 0 \\
-k' & 0 & 0 
\end{pmatrix}\,, \quad k'>0\,,  \quad\text{or}\\[2mm]
\label{eq:tex3}
M_q \,&:\,
 \begin{pmatrix}
0 & k' & k+k' \\
-k' & 0 & k \\
-k-k' & -k & 0 
\end{pmatrix}\,, \quad k,k' >0\,,
\end{align}
%
in both sectors ($k_{ij}^q=k_{ij}$), depending on the number of negative-weight entries in $M_q$ and,
in the case of 2 such entries,
on whether these share a row or a column.

One can then classify potentially viable models in terms of their number of parameters.
We denote $Q = Q_1 \oplus Q_2 \oplus Q_3$ and $q^c = q^c_1 \oplus q^c_2 \oplus q^c_3$ ($q=u,d$).
To avoid a massless quark, one must have
\begin{equation}
\begin{aligned}
Q &\sim (\mathbf{1}_{Q_1},k_1)\oplus  (\mathbf{1}_{Q_2},k_2)\oplus  (\mathbf{1}_{Q_3},k_3)\,,\\[1mm]
u^c &\sim (\overline{\mathbf{1}_{Q_1}},-k_1)\oplus  (\overline{\mathbf{1}_{Q_2}},-k_2)\oplus (\overline{\mathbf{1}_{Q_3}},-k_3)\,,\\[1mm]
d^c &\sim (\overline{\mathbf{1}_{Q_1}},-k_1)\oplus  (\overline{\mathbf{1}_{Q_2}},-k_2)\oplus (\overline{\mathbf{1}_{Q_3}},-k_3)\,,
\end{aligned}
\end{equation}
%
with $k' = k_{12} = k_1 - k_2$ and $k = k_{23} = k_2 - k_3$.

\subsection{Model landscape}
\label{sec:1Dtextures}

Given the above, there are five inequivalent textures one may consider for the $M_q$.
In an appropriate weak basis, they read, up to an overall factor $v_q$,%
\footnote{%
Even though textures I and II resemble those of~\cref{eq:2+1} in an appropriate basis, they are not automatically excluded (cf.~\cref{sec:2+1}) as they allow different values along the diagonal.}
\begin{subequations} \label[pluralequation]{eq:ts}
\begin{align} \label{eq:t1}
\text{I : }\,\,\,\; & \begin{pmatrix}
\alpha_1^q & 0 & \tilde\alpha_{13}^q\, \mathcal{Y}^{(k+k')}_{q,{13}}  \\[2mm]
0 & \alpha_2^q & \tilde\alpha_{23}^q \,\mathcal{Y}^{(k)}_{q,{23}}  \\[2mm]
\,\, 0 \,\,{} & \qquad 0 \qquad{} & \qquad \alpha_3^q \qquad{}
\end{pmatrix} \,,\,\, k>0,\, k'\geq 0\,,
 \\[2mm]\label{eq:t2}
\text{II :}\;\;\, & \begin{pmatrix}
\alpha_1^q & \tilde\alpha_{12}^q\, \mathcal{Y}^{(k')}_{q,12}  & \tilde\alpha_{13}^q\, \mathcal{Y}^{(k+k')}_{q,{13}}  \\[2mm]
0 & \alpha_2^q & 0 \\[2mm]
\,\, 0 \,\,{} & \qquad 0 \qquad{} & \qquad \alpha_3^q \qquad{}
\end{pmatrix}\,,\,\, k\geq 0,\, k'> 0\,,
 \\[2mm]\label{eq:t3}
\text{III :}\,\, & \begin{pmatrix}
\alpha_1^q &  \tilde\alpha_{12}^q\, \mathcal{Y}^{(k')}_{q,{12}} & \tilde\alpha_{13}^q\, \mathcal{Y}^{(k+k')}_{q,{13}}  \\[2mm]
0 & \alpha_2^q & \tilde\alpha_{23}^q \,\mathcal{Y}^{(k)}_{q,{23}}  \\[2mm]
\,\, 0 \,\,{} & \qquad 0 \qquad{} & \qquad \alpha_3^q \qquad{}
\end{pmatrix} \,,\,\, k,k'>0\,,
 \\[5mm]\label{eq:t4}
\text{IV :}\,\, & \begin{pmatrix}
\alpha_{11}^q &  \alpha_{12}^q & \tilde\alpha_{13}^q\, \mathcal{Y}^{(k)}_{q,{13}}  \\[2mm]
\alpha_{21}^q & \alpha_{22}^q & \tilde\alpha_{23}^q \,\mathcal{Y}^{(k)}_{q,{23}}  \\[2mm]
\,\, 0 \,\,{} & \hspace{17pt} 0 \hspace{17pt}{} &  \hspace{21.4pt} \alpha_3^q \hspace{21.4pt}{}
\end{pmatrix} \,,\,\, k>0\,,
 \\[5mm]\label{eq:t5}
\text{V :}\,\, & \begin{pmatrix}
\alpha_1^q &  \tilde\alpha_{12}^q\, \mathcal{Y}^{(k')}_{q,{12}} & \tilde\alpha_{13}^q\, \mathcal{Y}^{(k')}_{q,{13}}  \\[2mm]
0 & \alpha_{22}^q & \alpha_{23}^q  \\[2mm]
\,\, 0 \,\,{} &  \quad\,\,\, \alpha_{32}^q  \quad\,\,\,{} & \hspace{18.1pt} \alpha_{33}^q \hspace{18.1pt}{}
\end{pmatrix} \,,\,\, k'>0\,,
\end{align}
\end{subequations}
%
with
\begin{equation}
 \mathcal{Y}_{q,ij}^{(w)} \equiv \sum_{n=1} g_n^q\, Y_{\mathbf{1}_{ij},n}^{(w)}\,,
 \qquad
g_n^q \equiv \frac{\alpha^q_{ij,n}}{\alpha_{ij}^q}
\,,
\end{equation}
%
and $\alpha_{ij,1} \equiv \alpha_{ij}$. The sum in $n$  goes over all available independent 1D modular forms
of weight $w=k_{ij}$ and irrep $\mathbf{1}_{ij} \equiv \overline{\mathbf{1}_{Q_i}} \otimes \mathbf{1}_{Q_j}$.
Singlet modular form irreps available at a certain weight $w \leq 36$, for a given level $N \leq 5$, are listed in~\cref{tab:1forms} of~\cref{app:forms}.
The effects of canonical normalization have been included in~\cref{eq:t1,eq:t2,eq:t3,eq:t4,eq:t5} via the definitions
\begin{equation}
\tilde \alpha^q_{ij} \equiv \alpha^q_{ij}\, ({2\im\tau})^{k_{ij}/2} 
\,,
\end{equation}
%
where $\alpha^q_{ij}$ are the original superpotential parameters.
As before, the $g$ and $\mathcal{Y}$ remain unchanged by this normalization.

\vskip 2mm
The vanishing of the $(i,j) = (1,2)$ and $(2,3)$ entries in textures I and II, respectively,
can be consistently achieved in the following conditions:
\begin{itemize}
\item[i)] for $k_{ij}>0$: if there are no modular forms of the requisite irrep $\mathbf{1}_{ij}$ at the corresponding weight $k_{ij}$;
\item[ii)]for $k_{ij}=0$: if $\mathbf{1}_{ij} \neq \mathbf{1}$, guaranteeing also that the $(j,i)$ zero-weight entry vanishes, as $\mathbf{1}_{ji} = \overline{\mathbf{1}_{ij}} \neq \mathbf{1}$. 
\end{itemize}
No more entries may vanish, as this would lead either to a massless quark or to a vanishing \ac{CKM} angle.
In contrast, one cannot consistently forbid entry $(1,3)$ of texture III, since if forms $Y^{(k')}_{\mathbf{1}_{12}}$ and $Y^{(k)}_{\mathbf{1}_{23}}$ are available, their product is a modular form of the requisite weight $k+k'$ and irrep $\mathbf{1}_{13} = \mathbf{1}_{12} \otimes \mathbf{1}_{23}$.

\vskip 2mm
Note that by working with the finite modular groups $\Gamma_N'$ we can access non-trivial singlet irreps and find mass matrices solving the strong CP problem beyond those considered in the context of the full modular group~\cite{Feruglio:2023uof}. The key differences are:
\begin{itemize}
\item[i)] one now has access to modular forms $Y \sim \mathbf{1}^*$ aside from trivial singlets $\mathbf{1}$ (polynomials in $E_4$ and $E_6$);
\item[ii)] one can now forbid certain zero-weight entries of the mass matrices by appropriate choices of $\mathbf{1}_{Q_i}$.
\end{itemize}
It is this last possibility that allows access to textures I and II with vanishing $k'$ and $k$, respectively. It requires a non-trivial multiplier system for the matter fields --- an option alluded to in Appendix A of Ref.~\cite{Feruglio:2023uof} --- which, in the context of the finite modular groups, corresponds to assigning these fields to non-trivial 1D irreps of $\Gamma_N'$.

\vskip 2mm

Naïvely, the minimal scenarios would have \emph{only one} singlet modular form contributing to each non-zero positive-weight entry and would feature 5 (cases I, II), 6 (case III),  or 7 (cases IV, V) real parameters in each sector, not counting $\tau$.
However, these simple possibilities are excluded, as they conserve CP.
Indeed, recall that the textures for both up and down quark sectors coincide in an appropriate weak basis, as mass matrices have a common weight and irrep structure.
This leads to a situation analogous to the one discussed in~\cref{sec:2+1}. Namely, for cases I and II, sector-independent phases factorize and cancel in the determination of $U_\text{CKM}$.
This is also true for cases III--V. In case III, if only one form is available in each entry, the relevant phases are related by
\begin{equation}
\arg Y^{(k+k')}_{\mathbf{1}_{13}} = \arg Y^{(k')}_{\mathbf{1}_{12}} + \arg Y^{(k)}_{\mathbf{1}_{23}}
\,,
\end{equation}
%
leading to CP conservation.
In case IV, $k_{13} = k_{23}$ and the fact that the zero-weight $(1,2)$ entry is allowed by the symmetry implies $\mathbf{1}_{12} = \mathbf{1}$, forcing $\mathbf{1}_{Q_1} = \mathbf{1}_{Q_2}$ and thus $\mathbf{1}_{13} = \mathbf{1}_{23}$.%
\footnote{This is also why one cannot consistently allow only one non-zero positive-weight entry in texture IV (similarly for V).}
A similar reasoning applies to case V.
So, in these last two cases, if a single form contributes to each positive-weight entry, it must be the same form, implying that phases factorize and CP is conserved.
As a cross-check, one may also compute a CP-odd weak-basis invariant like $\tr[M_uM_u^\dagger, M_dM_d^\dagger]^3$~\cite{Bernabeu:1986fc} and verify that it vanishes when phases are common to both sectors.

The problem of CP conservation due to common phases can be avoided if more than one form is present in some entry of the mass matrices.
The \emph{minimal} matrices are then obtained from textures I and II of~\cref{eq:t1,eq:t2} whenever two singlet modular forms may contribute to one of the positive-weight entries, while the other positive-weight entry depends on a single modular form.
The minimal viable number of parameters thus corresponds to \emph{6 real constants in each sector}. Taking into account $\re \tau$ and $\im \tau$, this leads to a total of 14 parameters in the quark sector.%
\footnote{Note that the \ac{VEV} of $\tau$ may be shared with the lepton sector or fixed at a particular, motivated value in extended constructions.}
\emph{Next-to-minimal} matrices instead correspond to:
\begin{itemize}
\item textures I and II~\cref{eq:t1,eq:t2} if four independent modular forms are involved, and
\item texture III of~\cref{eq:t3} with two singlet modular forms contributing to one of the three positive-weight entries, while the other two depend on a single form each. By consistency, the entry populated with two independent forms must be $(1,3)$.
\end{itemize}
Such next-to-minimal scenarios feature \emph{7 real constants in each sector}, for a total of 16 parameters. 

\vskip 2mm

After going through the available forms and irreps (see~\cref{tab:1forms} of~\cref{app:forms}), we summarize all possible minimal and next-to-minimal models 
in~\cref{tab:models}. 
Each pair of numbers denotes the weights of the two lowest-weight modular forms involved, while primes and hats denote the corresponding singlet irreps. For instance, the $S_4'$-specific \mz{\hatp{7}'}{12} refers to the minimal textures
\begin{equation}
\begin{aligned}
  &\begin{pmatrix}
\alpha_1^q & 0 & \tilde\alpha_{13,1}^q\, Y^{(12)}_{\mathbf{1},1} + \tilde\alpha_{13,2}^q\, Y^{(12)}_{\mathbf{1},2}  \\[2mm]
0 & \alpha_2^q & \tilde\alpha_{23}^q \,Y^{(7)}_{\mathbf{\hat{1}'}}  \\[2mm]
\,\, 0 \,\,{} & \qquad 0 \qquad{} & \qquad \alpha_3^q \qquad{}
\end{pmatrix}\,\text{and}\\[2mm]
  &\begin{pmatrix}
\alpha_1^q & \tilde\alpha_{23}^q \,Y^{(7)}_{\mathbf{\hat{1}'}} & \tilde\alpha_{13,1}^q\, Y^{(12)}_{\mathbf{1},1} + \tilde\alpha_{13,2}^q\, Y^{(12)}_{\mathbf{1},2}  \\[2mm]
0 & \alpha_2^q & 0  \\[2mm]
\,\, 0 \,\,{} & \qquad 0 \qquad{} & \qquad \alpha_3^q \qquad{}
\end{pmatrix}
\end{aligned}
\end{equation}
%
of type I and II, respectively. Here and in what follows, the level $N$ is implied. Instead, the $A_4'$-specific \mz{4'}{8''} refers to the next-to-minimal textures
\begin{equation}
\begin{aligned}
  &\begin{pmatrix}
\alpha_1^q &  \tilde\alpha_{12}^q \,Y^{(8)}_{\mathbf{1''}}   & \tilde\alpha_{13,1}^q\, Y^{(12)}_{\mathbf{1},1} + \tilde\alpha_{13,2}^q\, Y^{(12)}_{\mathbf{1},2}  \\[2mm]
0 & \alpha_2^q & \tilde\alpha_{23}^q \,Y^{(4)}_{\mathbf{1'}}  \\[2mm]
\,\, 0 \,\,{} & \qquad 0 \qquad{} & \qquad \alpha_3^q \qquad{}
\end{pmatrix}\,\text{and}\\[2mm]
  &\begin{pmatrix}
\alpha_1^q &  \tilde\alpha_{12}^q \,Y^{(4)}_{\mathbf{1'}}   & \tilde\alpha_{13,1}^q\, Y^{(12)}_{\mathbf{1},1} + \tilde\alpha_{13,2}^q\, Y^{(12)}_{\mathbf{1},2}  \\[2mm]
0 & \alpha_2^q & \tilde\alpha_{23}^q \,Y^{(8)}_{\mathbf{1''}}  \\[2mm]
\,\, 0 \,\,{} & \qquad 0 \qquad{} & \qquad \alpha_3^q \qquad{}
\end{pmatrix}
\end{aligned}
\end{equation}
%
of type III. In general, each pair refers to two inequivalent models related by the transposition of mass matrices, as in the above examples. The exceptions are symmetric pairs arising for type-III next-to-minimal models, e.g.~\mz{6}{6}, which refer to a single texture.

We find a total of 462 minimal and 875 next-to-minimal inequivalent models.
Out of the minimal models, 6 are available for all the $\Gamma_N'$, as these can be realized using polynomials in $Y^{(4)}_{\mathbf{1}} = E_4$ and $Y^{(6)}_{\mathbf{1}} = E_6$, the (trivial) singlets of $SL(2,\mathbb{Z})$ common to all finite modular groups. There are additionally 48, 138 and 270 minimal models specific to $S_3$, $A_4'$ and $S_4'$, respectively ($A_5'$ does not have non-trivial singlets).
As for the next-to-minimal models of type I and II, 6 are available for all $\Gamma_N'$, while there are 48, 116 and 270 minimal models specific to $S_3$, $A_4'$ and $S_4'$, respectively. Out of these 440 models, 60 involve weights and irreps corresponding to three available independent modular forms.
Finally, there are 435 next-to-minimal models of type III, 15 of which are available for all $\Gamma_N'$,  while 48, 129 and 243 are specific to $S_3$, $A_4'$ and $S_4'$, respectively.
We stress that these lists are complete as no weight limit is being applied.%
\footnote{%
Only the next-to-minimal pairs \mz{4}{8}, \mz{6}{6}, \mz{6}{10}, and \mz{8}{8}, featuring 7 parameters per sector, and a model with 8 parameters per sector were considered in Ref.~\cite{Feruglio:2023uof}, whose selection is based on minimal field weights.}
Any other viable choice of weights will result in more than 7 parameters in each sector (at least 18 parameters in total).

\newcommand{\mI}[2]{\scriptsize \textcolor{black}{$(#1,#2)$}\vphantom{$\hatp{10}'$}}
\newcommand{\mtI}[3]{\scriptsize \textcolor{black}{$(#1,#2)$}\vphantom{$\hatp{10}'$}}

\begin{table*}[p]
  \centering
  \caption{Complete landscape of minimal and next-to-minimal models.
}
  \label{tab:models}
  \begin{tabular}{p{1cm}C{5.2cm}C{5.2cm}C{5.2cm}}
    \toprule
     & Minimal models (I and II) & Next-to-minimal models (I and II) & Next-to-minimal models (III) \\
    \midrule
All $\Gamma_N'$ 
& \mI{10}{12}, \mI{12}{14}, \mI{14}{16}
& \m{16}{18}, \m{18}{20}, \m{20}{22} 
& \mt{4}{8}{12}, \mt{4}{14}{18}, \mt{6}{6}{12}, \mt{6}{10}{16}, \mt{6}{14}{20}, \mt{8}{8}{16}, \mt{8}{10}{18}, \mt{8}{14}{22}, \mt{10}{10}{20}
\\[2mm]
$S_3$ only 
& \m{10'}{12}, \m{10}{18'}, \mI{10'}{18}, \m{12}{12'}, \m{12}{14'}, \m{12}{16'}, \mI{12'}{16}, \m{12}{20'}, \mI{12'}{20}, \mI{14'}{16}, \m{14}{18'}, \mI{14'}{18}, \m{14}{22'}, \mI{14'}{22}, \mI{16}{16'}, \mI{16'}{18}, \m{16'}{18'}, \mI{16}{20'}, \mI{16'}{20}, \mI{18}{20'}, \m{18'}{20'}, \mI{20}{20'}, \mI{20'}{22}, \m{20'}{22'}
& \m{16}{18'}, \m{16}{24'}, \m{16'}{24}, \m{18}{18'}, \m{18'}{20}, \m{18}{22'}, \m{18'}{22}, \m{18}{26'}, \m{18'}{26}, \m{20}{22'}, \m{20}{24'}, \m{20'}{24}, \m{20}{28'}, \m{20'}{28}, \m{22}{22'}, \m{22}{24'}, \m{22'}{24'}, \m{22}{26'}, \m{22'}{26}, \m{24'}{26}, \m{24'}{26'}, \m{26}{26'}, \m{26}{28'}, \m{26'}{28'}
& \mt{4}{14'}{18'}, \mt{4}{20'}{24'}, \mt{6'}{6'}{12}, \mt{6'}{10'}{16}, \mt{6}{12'}{18'}, \mt{6'}{12'}{18}, \mt{6'}{14'}{20}, \mt{6}{16'}{22'}, \mt{6'}{16'}{22}, \mt{6}{20'}{26'}, \mt{6'}{20'}{26}, \mt{8}{10'}{18'}, \mt{8}{14'}{22'}, \mt{8}{16'}{24'}, \mt{8}{20'}{28'}, \mt{10'}{10'}{20}, \mt{10}{12'}{22'}, \mt{10'}{12'}{22}, \mt{10}{14'}{24'}, \mt{10'}{14}{24'}, \mt{10}{16'}{26'}, \mt{10'}{16'}{26}, \mt{12'}{14}{26'}, \mt{12'}{14'}{26}, \mt{14}{14'}{28'}
\\[3mm]
$A_4'$ only 
& \m{8'}{12}, \mI{8'}{18}, \m{10'}{12}, \m{10}{16'}, \mI{10'}{16}, \m{10}{20''}, \mI{10'}{20}, \m{12}{12'}, \m{12}{12''}, \m{12}{14'}, \m{12}{14''}, \m{12}{16''}, \mI{12'}{16}, \m{12''}{16'}, \m{12}{18'}, \m{12}{18''}, \mI{12'}{18}, \mI{12''}{18}, \m{12}{22''}, \mI{12'}{22}, \m{12''}{22'}, \m{14}{16'}, \mI{14'}{16}, \m{14'}{16'}, \mI{14''}{16}, \m{14''}{16'}, \mI{14'}{18}, \m{14}{20'}, \m{14}{20''}, \mI{14'}{20}, \m{14'}{20''}, \mI{14''}{20}, \m{14''}{20'}, \m{14}{24''}, \m{14''}{24'}, \mI{16}{16''}, \m{16'}{16''}, \mI{16}{18'}, \mI{16}{18''}, \m{16'}{18'}, \m{16'}{18''}, \mI{16''}{18}, \m{16''}{20'}, \mI{16}{22''}, \m{16'}{22''}, \mI{16''}{22}, \m{16''}{22'}, \m{16''}{26'}, \mI{18}{18'}, \mI{18}{18''}, \mI{18'}{20}, \m{18'}{20'}, \m{18'}{20''}, \mI{18''}{20}, \m{18''}{20'}, \m{18''}{20''}, \mI{18}{22''}, \mI{18'}{22}, \m{18''}{22'}, \m{18'}{24''}, \m{18''}{24'}, \mI{20}{22''}, \m{20'}{22''}, \m{20''}{22''}, \mI{22}{22''}, \m{22'}{22''}, \m{22''}{24'}, \m{22''}{24''}, \m{22''}{26'}
& \m{14'}{24}, \m{16}{16'}, \m{16'}{18}, \m{16}{20''}, \m{16'}{20}, \m{16}{22'}, \m{16'}{22}, \m{16}{26''}, \m{16'}{26}, \m{18}{20'}, \m{18}{20''}, \m{18}{24'}, \m{18}{24''}, \m{18'}{24}, \m{18''}{24}, \m{18}{28''}, \m{18'}{28}, \m{18''}{28'}, \m{20}{20'}, \m{20}{20''}, \m{20'}{20''}, \m{20}{22'}, \m{20'}{22}, \m{20'}{22'}, \m{20''}{22}, \m{20''}{22'}, \m{20}{24''}, \m{20''}{24'}, \m{20}{26'}, \m{20}{26''}, \m{20'}{26}, \m{20'}{26''}, \m{20''}{26}, \m{20''}{26'}, \m{20}{30''}, \m{20''}{30'}, \m{22}{22'}, \m{22}{24'}, \m{22}{24''}, \m{22'}{24'}, \m{22'}{24''}, \m{22''}{24}, \m{22}{26''}, \m{22'}{26}, \m{22}{28''}, \m{22'}{28''}, \m{22''}{28}, \m{22''}{28'}, \m{22''}{32'}, \m{24'}{24''}, \m{24'}{26}, \m{24'}{26'}, \m{24'}{26''}, \m{24''}{26}, \m{24''}{26'}, \m{24''}{26''}, \m{24'}{30''}, \m{24''}{30'}
& \mt{4'}{8''}{12}, \mt{4}{12'}{16'}, \mt{4'}{12''}{16}, \mt{4'}{14''}{18}, \mt{4}{16''}{20''}, \mt{4'}{16''}{20}, \mt{4}{18'}{22'}, \mt{4'}{18''}{22}, \mt{4}{22''}{26''}, \mt{4'}{22''}{26}, \mt{6}{10'}{16'}, \mt{6}{14'}{20'}, \mt{6}{14''}{20''}, \mt{6}{18'}{24'}, \mt{6}{18''}{24''}, \mt{6}{22''}{28''}, \mt{8}{8'}{16'}, \mt{8'}{8''}{16}, \mt{8''}{8''}{16'}, \mt{8''}{10'}{18}, \mt{8}{12'}{20'}, \mt{8}{12''}{20''}, \mt{8'}{12'}{20''}, \mt{8'}{12''}{20}, \mt{8''}{12'}{20}, \mt{8''}{12''}{20'}, \mt{8}{14'}{22'}, \mt{8'}{14}{22'}, \mt{8'}{14''}{22}, \mt{8''}{14'}{22}, \mt{8''}{14''}{22'}, \mt{8}{16''}{24''}, \mt{8''}{16''}{24'}, \mt{8}{18'}{26'}, \mt{8}{18''}{26''}, \mt{8'}{18'}{26''}, \mt{8'}{18''}{26}, \mt{8''}{18'}{26}, \mt{8''}{18''}{26'}, \mt{8}{22''}{30''}, \mt{8''}{22''}{30'}, \mt{10}{10'}{20'}, \mt{10'}{10'}{20''}, \mt{10}{12'}{22'}, \mt{10'}{12''}{22}, \mt{10}{14'}{24'}, \mt{10}{14''}{24''}, \mt{10'}{14}{24'}, \mt{10'}{14'}{24''}, \mt{10}{16''}{26''}, \mt{10'}{16''}{26}, \mt{10}{18''}{28''}, \mt{10'}{18'}{28''}, \mt{12'}{12'}{24''}, \mt{12''}{12''}{24'}, \mt{12'}{14}{26'}, \mt{12'}{14'}{26''}, \mt{12'}{14''}{26}, \mt{12''}{14}{26''}, \mt{12''}{14'}{26}, \mt{12''}{14''}{26'}, \mt{12'}{18'}{30''}, \mt{12''}{18''}{30'}, \mt{14}{14''}{28''}, \mt{14'}{14'}{28''}, \mt{14}{16''}{30''}, \mt{14''}{16''}{30'}
\\[6mm]
$S_4'$ only 
& \m{\hatp{7}'}{12}, \mI{\hatp{7}'}{18}, \m{\hatp{9}'}{12}, \mI{\hatp{9}'}{16}, \mI{\hatp{9}'}{20}, \m{10'}{12}, \m{10}{\hatp{15}'}, \m{10'}{\hatp{15}'}, \m{10}{18'}, \mI{10'}{18}, \m{10}{\hatp{21}}, \m{10'}{\hatp{21}'}, \m{\hatp{11}'}{12}, \mI{\hatp{11}'}{16}, \mI{\hatp{11}'}{18}, \mI{\hatp{11}'}{22}, \m{12}{12'}, \m{12}{\hatp{13}}, \m{12}{\hatp{13}'}, \m{12}{14'}, \m{12}{\hatp{15}}, \m{12'}{\hatp{15}'}, \m{12}{16'}, \mI{12'}{16}, \m{12}{\hatp{17}}, \m{12}{\hatp{17}'}, \m{12}{\hatp{19}}, \m{12'}{\hatp{19}'}, \m{12}{20'}, \mI{12'}{20}, \m{12}{\hatp{23}}, \m{12'}{\hatp{23}'}, \m{\hatp{13}}{\hatp{15}'}, \m{\hatp{13}'}{\hatp{15}'}, \mI{\hatp{13}'}{16}, \mI{\hatp{13}}{18}, \m{\hatp{13}}{18'}, \mI{\hatp{13}'}{18}, \m{\hatp{13}'}{18'}, \mI{\hatp{13}'}{20}, \m{\hatp{13}}{\hatp{21}'}, \m{\hatp{13}'}{\hatp{21}}, \m{\hatp{13}}{24'}, \m{14}{\hatp{15}'}, \m{14'}{\hatp{15}'}, \mI{14'}{16}, \m{14}{18'}, \mI{14'}{18}, \m{14}{\hatp{19}'}, \m{14'}{\hatp{19}'}, \m{14}{\hatp{21}}, \m{14'}{\hatp{21}'}, \m{14}{22'}, \mI{14'}{22}, \m{14}{\hatp{25}}, \m{14'}{\hatp{25}'}, \m{\hatp{15}}{\hatp{15}'}, \mI{\hatp{15}}{16}, \m{\hatp{15}'}{16'}, \m{\hatp{15}'}{\hatp{17}}, \m{\hatp{15}'}{\hatp{17}'}, \m{\hatp{15}}{18'}, \m{\hatp{15}}{\hatp{19}'}, \m{\hatp{15}'}{\hatp{19}}, \mI{\hatp{15}}{20}, \m{\hatp{15}'}{20'}, \m{\hatp{15}}{22'}, \m{\hatp{15}}{\hatp{23}'}, \m{\hatp{15}'}{\hatp{23}}, \m{\hatp{15}}{26'}, \mI{16}{16'}, \mI{16}{\hatp{17}}, \m{16}{\hatp{17}'}, \mI{16'}{18}, \m{16'}{18'}, \mI{16}{\hatp{19}}, \m{16'}{\hatp{19}'}, \mI{16}{20'}, \mI{16'}{20}, \m{16'}{\hatp{21}}, \m{16'}{\hatp{21}'}, \mI{16}{\hatp{23}}, \m{16'}{\hatp{23}'}, \mI{\hatp{17}}{18}, \m{\hatp{17}}{18'}, \m{\hatp{17}'}{18}, \m{\hatp{17}'}{18'}, \m{\hatp{17}}{\hatp{19}'}, \m{\hatp{17}'}{\hatp{19}'}, \mI{\hatp{17}'}{20}, \m{\hatp{17}}{\hatp{21}'}, \m{\hatp{17}'}{\hatp{21}}, \mI{\hatp{17}}{22}, \m{\hatp{17}}{22'}, \m{\hatp{17}'}{22}, \m{\hatp{17}'}{22'}, \m{\hatp{17}}{24'}, \m{\hatp{17}}{\hatp{25}'}, \m{\hatp{17}'}{\hatp{25}}, \m{\hatp{17}}{28'}, \mI{18}{\hatp{19}}, \m{18'}{\hatp{19}}, \mI{18}{20'}, \m{18'}{20'}, \mI{18}{\hatp{23}}, \m{18'}{\hatp{23}}, \m{\hatp{19}}{\hatp{19}'}, \mI{\hatp{19}}{20}, \m{\hatp{19}'}{20'}, \m{\hatp{19}}{\hatp{21}}, \m{\hatp{19}}{\hatp{21}'}, \m{\hatp{19}}{22'}, \m{\hatp{19}}{\hatp{23}'}, \m{\hatp{19}'}{\hatp{23}}, \m{\hatp{19}}{24'}, \m{\hatp{19}}{26'}, \mI{20}{20'}, \m{20'}{\hatp{21}}, \m{20'}{\hatp{21}'}, \mI{20'}{22}, \m{20'}{22'}, \mI{20}{\hatp{23}}, \m{20'}{\hatp{23}'}, \m{20'}{\hatp{25}}, \m{20'}{\hatp{25}'}, \m{\hatp{21}}{\hatp{23}}, \m{\hatp{21}'}{\hatp{23}}, \mI{22}{\hatp{23}}, \m{22'}{\hatp{23}}, \m{\hatp{23}}{\hatp{23}'}, \m{\hatp{23}}{24'}, \m{\hatp{23}}{\hatp{25}}, \m{\hatp{23}}{\hatp{25}'}, \m{\hatp{23}}{26'}, \m{\hatp{23}}{28'}
& \m{\hatp{13}'}{24}, \m{\hatp{15}'}{16}, \m{\hatp{15}'}{18}, \m{\hatp{15}'}{20}, \m{\hatp{15}'}{22}, \m{\hatp{15}'}{26}, \m{16}{18'}, \m{16}{\hatp{21}}, \m{16}{\hatp{21}'}, \m{16}{24'}, \m{16'}{24}, \m{16}{\hatp{27}}, \m{16'}{\hatp{27}'}, \m{\hatp{17}'}{24}, \m{\hatp{17}'}{28}, \m{18}{18'}, \m{18}{\hatp{19}'}, \m{18'}{\hatp{19}'}, \m{18'}{20}, \m{18}{\hatp{21}}, \m{18'}{\hatp{21}'}, \m{18}{22'}, \m{18'}{22}, \m{18}{\hatp{23}'}, \m{18'}{\hatp{23}'}, \m{18}{\hatp{25}}, \m{18'}{\hatp{25}'}, \m{18}{26'}, \m{18'}{26}, \m{18}{\hatp{29}}, \m{18'}{\hatp{29}'}, \m{\hatp{19}'}{20}, \m{\hatp{19}'}{\hatp{21}}, \m{\hatp{19}'}{\hatp{21}'}, \m{\hatp{19}'}{22}, \m{\hatp{19}}{24}, \m{\hatp{19}'}{24'}, \m{\hatp{19}'}{26}, \m{\hatp{19}}{\hatp{27}'}, \m{\hatp{19}'}{\hatp{27}}, \m{\hatp{19}}{30'}, \m{20}{\hatp{21}}, \m{20}{\hatp{21}'}, \m{20}{22'}, \m{20}{24'}, \m{20'}{24}, \m{20}{\hatp{25}}, \m{20}{\hatp{25}'}, \m{20}{\hatp{27}}, \m{20'}{\hatp{27}'}, \m{20}{28'}, \m{20'}{28}, \m{20}{\hatp{31}}, \m{20'}{\hatp{31}'}, \m{\hatp{21}}{\hatp{21}'}, \m{\hatp{21}}{22}, \m{\hatp{21}}{22'}, \m{\hatp{21}'}{22}, \m{\hatp{21}'}{22'}, \m{\hatp{21}}{\hatp{23}'}, \m{\hatp{21}'}{\hatp{23}'}, \m{\hatp{21}}{24'}, \m{\hatp{21}}{\hatp{25}'}, \m{\hatp{21}'}{\hatp{25}}, \m{\hatp{21}}{26}, \m{\hatp{21}}{26'}, \m{\hatp{21}'}{26}, \m{\hatp{21}'}{26'}, \m{\hatp{21}}{28'}, \m{\hatp{21}}{\hatp{29}'}, \m{\hatp{21}'}{\hatp{29}}, \m{\hatp{21}}{32'}, \m{22}{22'}, \m{22}{\hatp{23}'}, \m{22'}{\hatp{23}'}, \m{22}{24'}, \m{22'}{24'}, \m{22}{\hatp{25}}, \m{22'}{\hatp{25}'}, \m{22}{26'}, \m{22'}{26}, \m{22}{\hatp{27}}, \m{22'}{\hatp{27}}, \m{22}{\hatp{29}}, \m{22'}{\hatp{29}'}, \m{\hatp{23}}{24}, \m{\hatp{23}'}{24'}, \m{\hatp{23}'}{\hatp{25}}, \m{\hatp{23}'}{\hatp{25}'}, \m{\hatp{23}'}{26}, \m{\hatp{23}}{\hatp{27}'}, \m{\hatp{23}'}{\hatp{27}}, \m{\hatp{23}}{28}, \m{\hatp{23}'}{28'}, \m{\hatp{23}}{30'}, \m{\hatp{23}}{\hatp{31}'}, \m{\hatp{23}'}{\hatp{31}}, \m{\hatp{23}}{34'}, \m{24'}{\hatp{25}}, \m{24'}{\hatp{25}'}, \m{24'}{26}, \m{24'}{26'}, \m{24'}{\hatp{29}}, \m{24'}{\hatp{29}'}, \m{\hatp{25}}{\hatp{25}'}, \m{\hatp{25}}{26}, \m{\hatp{25}}{26'}, \m{\hatp{25}'}{26}, \m{\hatp{25}'}{26'}, \m{\hatp{25}}{\hatp{27}}, \m{\hatp{25}'}{\hatp{27}}, \m{\hatp{25}}{28'}, \m{\hatp{25}}{\hatp{29}'}, \m{\hatp{25}'}{\hatp{29}}, \m{\hatp{25}}{32'}, \m{26}{26'}, \m{26}{\hatp{27}}, \m{26'}{\hatp{27}}, \m{26}{28'}, \m{26'}{28'}, \m{26}{\hatp{29}}, \m{26'}{\hatp{29}'}, \m{26}{\hatp{31}}, \m{26'}{\hatp{31}}, \m{\hatp{27}}{28'}, \m{\hatp{27}}{\hatp{29}}, \m{\hatp{27}}{\hatp{29}'}, \m{\hatp{27}}{32'}, \m{28'}{\hatp{29}}, \m{28'}{\hatp{29}'}, \m{\hatp{29}}{\hatp{29}'}, \m{\hatp{29}}{\hatp{31}}, \m{\hatp{29}'}{\hatp{31}}, \m{\hatp{29}}{32'}, \m{\hatp{31}}{32'}
& \mt{\hatp{3}'}{\hatp{9}}{12}, \mt{\hatp{3}'}{\hatp{13}}{16}, \mt{\hatp{3}'}{\hatp{15}}{18}, \mt{\hatp{3}'}{\hatp{17}}{20}, \mt{\hatp{3}'}{\hatp{19}}{22}, \mt{\hatp{3}'}{\hatp{23}}{26}, \mt{4}{\hatp{11}'}{\hatp{15}'}, \mt{4}{14'}{18'}, \mt{4}{\hatp{17}}{\hatp{21}}, \mt{4}{\hatp{17}'}{\hatp{21}'}, \mt{4}{20'}{24'}, \mt{4}{\hatp{23}}{\hatp{27}}, \mt{6'}{6'}{12}, \mt{6}{\hatp{9}'}{\hatp{15}'}, \mt{6'}{\hatp{9}}{\hatp{15}'}, \mt{6'}{10'}{16}, \mt{6}{12'}{18'}, \mt{6'}{12'}{18}, \mt{6}{\hatp{13}'}{\hatp{19}'}, \mt{6'}{\hatp{13}}{\hatp{19}'}, \mt{6'}{14'}{20}, \mt{6}{\hatp{15}}{\hatp{21}}, \mt{6'}{\hatp{15}}{\hatp{21}'}, \mt{6}{16'}{22'}, \mt{6'}{16'}{22}, \mt{6}{\hatp{17}'}{\hatp{23}'}, \mt{6'}{\hatp{17}}{\hatp{23}'}, \mt{6}{\hatp{19}}{\hatp{25}}, \mt{6'}{\hatp{19}}{\hatp{25}'}, \mt{6}{20'}{26'}, \mt{6'}{20'}{26}, \mt{6}{\hatp{23}}{\hatp{29}}, \mt{6'}{\hatp{23}}{\hatp{29}'}, \mt{\hatp{7}'}{8}{\hatp{15}'}, \mt{\hatp{7}'}{\hatp{9}}{16}, \mt{\hatp{7}'}{\hatp{11}'}{18'}, \mt{\hatp{7}'}{\hatp{13}}{20}, \mt{\hatp{7}'}{14}{\hatp{21}'}, \mt{\hatp{7}'}{14'}{\hatp{21}}, \mt{\hatp{7}'}{\hatp{15}}{22}, \mt{\hatp{7}'}{\hatp{17}'}{24'}, \mt{\hatp{7}'}{\hatp{19}}{26}, \mt{\hatp{7}'}{20'}{\hatp{27}}, \mt{8}{10'}{18'}, \mt{8}{\hatp{11}'}{\hatp{19}'}, \mt{8}{\hatp{13}}{\hatp{21}}, \mt{8}{\hatp{13}'}{\hatp{21}'}, \mt{8}{14'}{22'}, \mt{8}{16'}{24'}, \mt{8}{\hatp{17}}{\hatp{25}}, \mt{8}{\hatp{17}'}{\hatp{25}'}, \mt{8}{\hatp{19}}{\hatp{27}}, \mt{8}{20'}{28'}, \mt{8}{\hatp{23}}{\hatp{31}}, \mt{\hatp{9}}{\hatp{9}}{18'}, \mt{\hatp{9}}{\hatp{9}'}{18}, \mt{\hatp{9}'}{\hatp{9}'}{18'}, \mt{\hatp{9}}{10'}{\hatp{19}'}, \mt{\hatp{9}'}{10}{\hatp{19}'}, \mt{\hatp{9}}{\hatp{11}'}{20}, \mt{\hatp{9}}{12'}{\hatp{21}'}, \mt{\hatp{9}'}{12'}{\hatp{21}}, \mt{\hatp{9}}{\hatp{13}}{22'}, \mt{\hatp{9}}{\hatp{13}'}{22}, \mt{\hatp{9}'}{\hatp{13}}{22}, \mt{\hatp{9}'}{\hatp{13}'}{22'}, \mt{\hatp{9}}{14'}{\hatp{23}'}, \mt{\hatp{9}'}{14}{\hatp{23}'}, \mt{\hatp{9}}{\hatp{15}}{24'}, \mt{\hatp{9}}{16'}{\hatp{25}'}, \mt{\hatp{9}'}{16'}{\hatp{25}}, \mt{\hatp{9}}{\hatp{17}}{26'}, \mt{\hatp{9}}{\hatp{17}'}{26}, \mt{\hatp{9}'}{\hatp{17}}{26}, \mt{\hatp{9}'}{\hatp{17}'}{26'}, \mt{\hatp{9}}{\hatp{19}}{28'}, \mt{\hatp{9}}{20'}{\hatp{29}'}, \mt{\hatp{9}'}{20'}{\hatp{29}}, \mt{\hatp{9}}{\hatp{23}}{32'}, \mt{10'}{10'}{20}, \mt{10}{\hatp{11}'}{\hatp{21}'}, \mt{10'}{\hatp{11}'}{\hatp{21}}, \mt{10}{12'}{22'}, \mt{10'}{12'}{22}, \mt{10}{\hatp{13}'}{\hatp{23}'}, \mt{10'}{\hatp{13}}{\hatp{23}'}, \mt{10}{14'}{24'}, \mt{10'}{14}{24'}, \mt{10}{\hatp{15}}{\hatp{25}}, \mt{10'}{\hatp{15}}{\hatp{25}'}, \mt{10}{16'}{26'}, \mt{10'}{16'}{26}, \mt{10}{\hatp{17}}{\hatp{27}}, \mt{10'}{\hatp{17}'}{\hatp{27}}, \mt{10}{\hatp{19}}{\hatp{29}}, \mt{10'}{\hatp{19}}{\hatp{29}'}, \mt{\hatp{11}'}{\hatp{11}'}{22'}, \mt{\hatp{11}'}{\hatp{13}'}{24'}, \mt{\hatp{11}'}{14}{\hatp{25}'}, \mt{\hatp{11}'}{14'}{\hatp{25}}, \mt{\hatp{11}'}{\hatp{15}}{26}, \mt{\hatp{11}'}{16'}{\hatp{27}}, \mt{\hatp{11}'}{\hatp{17}'}{28'}, \mt{\hatp{11}'}{20'}{\hatp{31}}, \mt{12'}{\hatp{13}}{\hatp{25}'}, \mt{12'}{\hatp{13}'}{\hatp{25}}, \mt{12'}{14}{26'}, \mt{12'}{14'}{26}, \mt{12'}{\hatp{17}}{\hatp{29}'}, \mt{12'}{\hatp{17}'}{\hatp{29}}, \mt{\hatp{13}}{\hatp{13}}{26'}, \mt{\hatp{13}}{\hatp{13}'}{26}, \mt{\hatp{13}'}{\hatp{13}'}{26'}, \mt{\hatp{13}}{14}{\hatp{27}}, \mt{\hatp{13}'}{14'}{\hatp{27}}, \mt{\hatp{13}}{\hatp{15}}{28'}, \mt{\hatp{13}}{16'}{\hatp{29}'}, \mt{\hatp{13}'}{16'}{\hatp{29}}, \mt{\hatp{13}}{\hatp{19}}{32'}, \mt{14}{14'}{28'}, \mt{14}{\hatp{15}}{\hatp{29}}, \mt{14'}{\hatp{15}}{\hatp{29}'}, \mt{14}{\hatp{17}}{\hatp{31}}, \mt{14'}{\hatp{17}'}{\hatp{31}}, \mt{\hatp{15}}{\hatp{17}}{32'}
\\
    \bottomrule
  \end{tabular}
\end{table*}
%

\subsection{Scan of minimal models}
\label{sec:scans}

Focusing on the minimal model landscape, we aim at confronting models with quark data, namely GUT-scale mass ratios, mixing angles and \ac{CPV} phase~\cite{deMedeirosVarzielas:2023crv}.
To this end, we start by fitting the textures themselves to data, i.e.~we consider textures I and II of \cref{eq:t1,eq:t2}
with their non-zero off-diagonal ($\tau$-dependent) entries replaced by free complex parameters.
Thanks to weak basis transformations, it turns out that only one complex phase is relevant for the scan.
Textures are fitted up to an overall factor in each sector, which dictates the up and down quark mass scales but does not affect mass ratios.
Following the procedure outlined in Appendix C of Ref.~\cite{Novichkov:2018ovf}, we first search for local minima of the Gaussian loss function and then explore $3\sigma$ viable regions around them via the Metropolis algorithm.

We find that both textures I and II provide enough freedom to fit the central values of quark mass ratios and mixing parameters, resulting in a negligible loss. This means that all models in~\cref{tab:models} --- including type-III next-to-minimal ones --- can fit the data, since texture III reduces to either texture I or texture II in the limit of vanishing $\alpha_{12,n}^q$ or $\alpha_{23,n}^q$, respectively.%
\footnote{The detailed analysis of non-minimal models is beyond the scope of this work.}
However, the fit may come at the price of extreme fine-tuning. Indeed, in the context of each model, each fit point may translate to severely hierarchical superpotential parameters.

To quantify the hierarchies among model parameters, we take the total logarithmic amplitude $\text{amp}_\text{tot} = \text{amp}_u + \text{amp}_d$, with
\begin{equation}
\text{amp}_q \equiv \max\, \log_{10} | \alpha^q| - \min\, \log_{10} | \alpha^q|\,,
\end{equation}
%
as a measure of model quality. By definition, $\text{amp}_q$ is small when all superpotential constants are of the same order in that sector, e.g.~if they are all $\mathcal{O}(1)$. A value of $\text{amp}_q = 1$ corresponds to $q$-sector constants differing at most by one order of magnitude among themselves.

In fitting textures I and II, the zero-weight (diagonal) entries alone already set a lower bound on the total amplitude, $\text{amp}_\text{tot} \geq \text{amp}_{\text{tot},0}$, for each fit point. Numerically, we find
\begin{equation} \label{eq:min0}
\min\,\text{amp}_{\text{tot},0}^\text{I} \simeq 1.99\,, \quad\,\,
\min\,\text{amp}_{\text{tot},0}^\text{II} \simeq 1.04 \,,
\end{equation}
%
using $\sim 2\times 10^6$ fit points for each texture.
This means that type-I models in general carry a stronger hierarchy in superpotential parameters from the outset, when compared to type-II models.

To further analyse the minimal models, we randomly sample the parameter space of each fit. In particular, we select a manageable sample of 1000 fit points for each texture, appending also the points selected in~\cref{eq:min0}. We then (exactly) fit each of the corresponding 231 minimal models to each sample point. Not only is this second fit always possible; it is degenerate in some parameters, as the models possess more freedom than the textures. We thus vary model parameters, including $\tau$, in order to minimize the amplitude \emph{excess}, defined as $\text{amp}_\text{tot} - \text{amp}_{\text{tot},0}$.
For this procedure to be meaningful, the normalization of modular forms must be appropriately fixed. The list of normalized singlet modular forms we use, based on the proposal of Ref.~\cite{Petcov:2023fwh}, is explicitly given in~\cref{app:forms}.

As a result of these scans, we identify the minimal models which can fit quark data with the lowest total amplitude:
\begin{itemize}
\item for texture I, the lowest-amplitude model is \mz{12}{14}, available for all $\Gamma_N'$ and in the context of the full modular group, with $\text{amp}_{\text{tot}} \simeq 2.19$ ($\text{amp}_{\text{tot},0} \simeq 1.99$, $\text{amp}_u \simeq 0.18 $, $\text{amp}_d \simeq 2.01$),
\item for texture II, the lowest-amplitude model is the $A_4^{(\prime)}$-exclusive \mz{16''}{22}, with $\text{amp}_{\text{tot}} \simeq 2.34$ ($\text{amp}_{\text{tot},0} \simeq 1.49$, $\text{amp}_u \simeq 0.05 $, $\text{amp}_d \simeq 2.29$).
\end{itemize}
Across low-amplitude minimal models, one finds an overall tendency for hierarchies to originate mostly from the down-quark sector, with $\text{amp}_d \sim 2$ vs.~$\text{amp}_u < 1$. 

\begin{figure*}[!htbp]
\centering
\includegraphics[width=1.1\columnwidth]{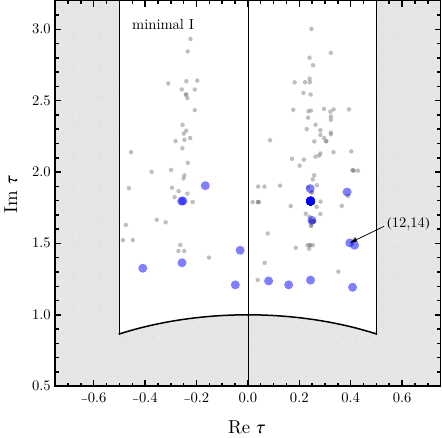}
\caption{Values of $\tau$ in the fundamental domain selected by minimizing the total amplitude for each type-I minimal model. 52 zero-excess models are not shown. Points shown in blue (grey) correspond to $\text{amp}_{\text{tot}} < 3$ ($\text{amp}_{\text{tot}}  >3$). The lowest-amplitude model is highlighted.}
\label{fig:tauI}
\vskip 1cm
\centering
\includegraphics[width=1.1\columnwidth]{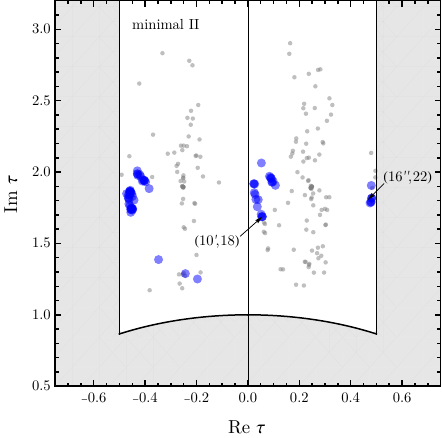}
\caption{The same as in~\cref{fig:tauI}, but for type-II minimal models. All 231 models minimize the total amplitude with non-zero excess. Both the lowest-amplitude model and the benchmark model of~\cref{sec:benchmark} are highlighted.}
\label{fig:tauII}
\end{figure*}

One may also wonder if special values of $\tau$ are selected by minimizing the total amplitude, i.e.~by asking for the smallest possible parametric hierarchies, in the context of each model. Note that an unambiguous selection of $\tau$ is not possible when the excess is zero.
In~\cref{fig:tauI,fig:tauII} we show the selected values of the modulus \ac{VEV} for type-I and type-II models with minimal $\text{amp}_\text{tot}$ and non-zero excess.
In the context of each model, any preference for either positive or negative values of $\re \tau$
must be driven by the \ac{CPV} phase $\delta$, which is the only observable sensitive to the sign of $\re \tau$ (see e.g.~\cite{Novichkov:2018ovf}).
Focusing on the models with relatively small total amplitude ($\text{amp}_{\text{tot}} < 3$, blue dots in the figures), we find 20 type-I models selecting $\tau \simeq 0.24 + 1.80\,i$ and no other apparent clusters. 
Instead, most low-amplitude type-II models cluster around $\tau \sim -0.45 + 1.8\,i$ (58\%), $\tau \sim 0.1 + 1.9\,i$ (29\%) or  $\tau \sim 0.45 + 1.8\,i$ (8\%).
To close, we present the details of a type-II model fit featuring a value of the modulus \ac{VEV} within the central cluster.

%

\subsection{An \texorpdfstring{$S_4$}{S4} benchmark}
\label{sec:benchmark}

As a benchmark model, we consider the $S_4^{(\prime)}$-specific type-II minimal model denoted by $(10',18)$. This modular model may arise as the result of, e.g., the following irrep and weight assignments under $\Gamma_4$:
\begin{equation}
\begin{aligned}
Q &\sim (\mathbf{1},9)\oplus  (\mathbf{1}',-1)\oplus  (\mathbf{1},-9)\,,\\[1mm]
u^c,\, d^c &\sim (\mathbf{1},-9)\oplus  (\mathbf{1}',1)\oplus (\mathbf{1},9)\,,
\end{aligned}
\end{equation}
%
resulting in 
$(k_{12}, \mathbf{1}_{12}) = (10,\mathbf{1}')$,
$(k_{13}, \mathbf{1}_{13}) = (18,\mathbf{1})$, and
$(k_{23}, \mathbf{1}_{23}) = (8,\mathbf{1}')$. The fact that there is no $\mathbf{1}'$ modular form of weight 8
guarantees the vanishing of the $(2,3)$ entries in both mass matrices $M_q$.

We present the found fit point for this model minimizing the total logarithmic amplitude, which was obtained for $\tau \simeq 0.06+1.69\,i$, as indicated in~\cref{fig:tauII}. The values of superpotential parameters are (cf.~\cref{eq:t2})
\begin{equation} \label{eq:alphas}
\begin{aligned}
&\alpha_2^u \simeq 1.00\, \alpha_1^u \,, \quad
\alpha_3^u \simeq 1.65\, \alpha_1^u \,, \quad
\alpha_{12}^u \simeq 1.27\, \alpha_1^u \,,\\[1mm]
&\alpha_{13,1}^u \simeq 1.47\, \alpha_1^u \,, \quad
\alpha_{13,2}^u \simeq 1.61\, \alpha_1^u \,,\\[3mm]
&\alpha_2^d \simeq 0.48\, \alpha_1^d \,, \quad
\alpha_3^d \simeq 0.05\, \alpha_1^d \,, \quad
\alpha_{12}^d \simeq 7.92\, \alpha_1^d \,,\\[1mm]
&\alpha_{13,1}^d \simeq -7.90\, \alpha_1^d \,, \quad
\alpha_{13,2}^d \simeq 0.05\, \alpha_1^d \,,
\end{aligned}
\end{equation}
%
leading to the logarithmic amplitudes
\begin{equation}
\begin{array}{l@{\qquad}l}
\text{amp}_{\text{tot}} \simeq 2.46\,, &
 \text{amp}_{\text{tot},0} \simeq 1.56\,, \\[2mm]
\text{amp}_u \simeq 0.22\,, &
\text{amp}_d \simeq 2.24\,.
\end{array}
\end{equation}
%
These values signal the fact that the hierarchy budget of the model is mostly associated with down sector constants, as can be checked from~\cref{eq:alphas}.
The resulting values for the observables are
\begin{equation}
\begin{aligned}
&m_u/m_c \simeq 4.41 \times 10^{-3}\,, \quad
m_c/m_t \simeq 2.67 \times 10^{-3}\,, \\
&m_d/m_s \simeq 5.61 \times 10^{-2}\,, \quad
m_s/m_b \simeq 1.45 \times 10^{-2}\,, \\[2mm]
&\theta_{12}^q \simeq 13.2\degree,\,\,
\theta_{23}^q \simeq 2.10\degree, \,\,
\theta_{13}^q \simeq 0.20\degree, \,\,
\delta \simeq 71.2\degree,
\end{aligned}
\end{equation}
%
corresponding to a successful fit of GUT-scale quark data at the $2.9 \sigma$ level.

\section{Summary and conclusions}
\label{sec:summary}

Modular symmetry can be used as a powerful generalization of traditional flavour symmetries. It may shed light on the observed patterns of fundamental fermion mixing and clarify the origins of fermion mass hierarchies and \acf{CPV}. Moreover, it has been shown that it may address the apparent absence of \ac{CPV} in the strong sector, i.e.~resolve the strong CP problem with high quality~\cite{Feruglio:2023uof}.

Here, we have identified the most general structures of modular weights (summarized in~\cref{eq:Mud}) which allow for a large \ac{CKM} phase $\delta$ while ensuring the vanishing of the QCD angle $\bar \theta$.
The crucial condition of anomaly cancellation --- implying zero modular weights for mass matrix determinants --- together with the requirement of no massless quarks, severely restricts such structures.
Note that this mechanism is, in essence, incompatible with the flavon-free mechanism of generating fermion mass hierarchies~\cite{Novichkov:2021evw}, as the latter requires non-zero weights for mass matrix determinants.

We have also generalized the existing approach, based on using the full modular group, to the finite modular groups $\Gamma_N'$. In doing so, one naturally gains access to i) modular forms aside from trivial singlets and to ii) the possibility of forbidding certain (zero-weight) entries of the mass matrices by appropriate choices of quark irreps.

A priori, one expects it may be possible to build viable models in this context involving quark irreps of dimension higher than one.  We show this is not the case. Namely, the only potentially viable scenario of this kind turns out to be the one  where all quark fields split into a doublet-plus-singlet configuration (see \cref{sec:2+1}). This setup is amenable to analytic treatment, which however reveals an unavoidable conflict with quark data, either with the values of $|V_{us}|$ and $|V_{ub}|$ or with $|V_{cb}|$.%
\footnote{These results are expected to hold even in extensions based on vector-valued modular forms.}
Hence, only scenarios where quarks furnish 1D irreps of the finite modular group have a chance to be phenomenologically viable.

In~\cref{sec:1+1+1}, we classify the potentially viable (singlet-based) quark mass structures, finding 5 inequivalent textures, see~\cref{eq:ts}.
We then focus on the models which feature a minimal or next-to-minimal number of real superpotential parameters, 12 and 14 respectively. These models are systematically identified and the corresponding landscape is presented in~\cref{tab:models}. 
Finally, we numerically scan the minimal model landscape. While there is enough freedom to fit the data, this may require hierarchical superpotential parameters. Asking for the smallest possible hierarchies reveals some preferences for certain regions within the fundamental modular domain (see~\cref{fig:tauI,fig:tauII}). An explicit $S_4$-based benchmark is presented in~\cref{sec:benchmark}.

At present, finding restrictive but viable modular-symmetric models of quark masses and mixing which naturally address both the strong CP problem and the origin of quark mass hierarchies and \ac{CPV} remains an interesting open challenge.

\vskip 5mm

\noindent
\emph{Note added:} While the numerical scan was being finalized,~\cite{Petcov:2024vph} appeared on the arXiv, exploring in detail one of the two $A_4^{(\prime)}$-exclusive (type-III) next-to-minimal models denoted here by $(4,12')$, cf.~\cref{tab:models}.

\section*{Acknowledgements}

The work of S.T.P.~was supported in part by the European Union's Horizon 2020 research and innovation programme under the Marie Sk\l{}odowska-Curie grant agreement No.~860881-HIDDeN, by the Italian INFN program on Theoretical Astroparticle Physics and by the World Premier International Research Center Initiative (WPI Initiative, MEXT), Japan.

\clearpage

\appendix
\section{Modular transformation of determinants}
\label{app:det}

Here we show explicitly that the determinant of a mass matrix $M$ in modular constructions is a singlet (i.e.~a \mbox{1-dimensional}) modular form of weight $\kdet = \sum_i k_i + k^c_i$. 
We start by noticing that, to guarantee modular invariance, the elements of the mass 
matrix $M$ --- which are obtained from modular forms --- must, 
in general, transform in the \ac{LR} convention we employ as (cf.~\cref{eq:modtrans}):
\begin{equation} \label{eq:Mij}
\begin{aligned}
M_{ij}(\tau) \,\xrightarrow{\gamma}\, &\,M_{ij}(\gamma\tau) \\
\,=\, &\,(c\tau + d)^{k_i + k_j^c} \,\rho_{ik}^*(\gamma)\, M_{kl}(\tau)\,\rho^{c\ast}_{jl}(\gamma) \,.
\end{aligned}
\end{equation}
%
We can now look into how the determinant of the mass matrix changes under a modular transformation:
\begin{widetext}
\begin{equation}
\begin{aligned}
\det M \,&=\, \frac{1}{n!}\, \varepsilon_{i_1\ldots i_n}\, \varepsilon_{j_1\ldots j_n}\, M_{i_1 j_1} \ldots M_{i_n j_n} \\[2mm]
\,&\xrightarrow{\gamma}\,
\frac{1}{n!}\, \varepsilon_{i_1\ldots i_n}\, \varepsilon_{j_1\ldots j_n}\, 
(c\tau + d)^{k_{i_1} + k^c_{j_1}} \rho^*_{i_1 k_1} \rho^{c\ast}_{j_1 l_1} M_{k_1 l_1} 
\\
&\hspace{3.2cm}\ldots \\
&\hspace{3.2cm}(c\tau + d)^{k_{i_n} + k^c_{j_n}} \rho^*_{i_n k_n} \rho^{c\ast}_{j_n l_n} M_{k_n l_n} 
\\[2mm]
\,&=\,
(c\tau + d)^{\sum_i k_i + k^c_i}
\frac{1}{n!}  
\left(\varepsilon_{i_1\ldots i_n}\,\rho^*_{i_1 k_1}\ldots \rho^*_{i_n k_n}\right)
\left(\varepsilon_{j_1\ldots j_n}\,\rho^{c\ast}_{j_1 l_1}\ldots \rho^{c\ast}_{j_n l_n}\right)
M_{k_1 l_1} \ldots M_{k_n l_n}
\\[2mm]
\,&=\,
(c\tau + d)^{\sum_i k_i + k^c_i}
\frac{1}{n!}  
\left(\det \rho^*\, \varepsilon_{k_1\ldots k_n}\right)
\left(\det \rho^{c\ast}\, \varepsilon_{l_1\ldots l_n}\right)
M_{k_1 l_1} \ldots M_{k_n l_n}
\\[2mm]
\,&=\,
(c\tau + d)^{\kdet}
\,\underbrace{(\det \rho)^*
\,(\det \rho^c)^*}_{\gamma\text{-dependent phase}}
\,\det M\,,
\end{aligned}
\end{equation}
%
where in the last line we have defined $\kdet \equiv \sum_i k_i + k^c_i$.
Alternatively, one can reach the same result by noticing that
$M(\gamma\tau)$ can be cast in the form
\begin{equation} \label{eq:Mgt}
M(\gamma\tau) \,=\, 
\begin{pmatrix}
(c\tau + d)^{k_1} & & \\
 & \ddots & \\
 & & (c\tau + d)^{k_n}
\end{pmatrix}
\rho^*(\gamma)\, M(\tau)\, \rho^{c\dagger}(\gamma) 
\begin{pmatrix}
(c\tau + d)^{k_1^c} & & \\
 & \ddots & \\
 & & (c\tau + d)^{k_n^c}
\end{pmatrix}\,,
\end{equation}
%
and by taking a determinant. Since the determinant of a product of matrices equals the 
product of their determinants, one finds $\det M (\gamma\tau) = 
(c\tau + d)^{\kdet}
\,(\det \rho)^*
\,(\det \rho^c)^*
\,\det M(\tau)$ directly.
\end{widetext}

Having found this transformation property, we verify that $\det M$ is a singlet modular form of weight $\kdet$. Note that it need not be a trivial singlet of the finite modular group, given the presence of a $\gamma$-dependent phase factor.
Indeed, consider as an example the mass matrix $M_e^\dagger$ of Ref.~\cite{Novichkov:2020eep}, obtained from level 4 modular forms, whose determinant is given in equation (E.4) therein. One can check then that, under $\tau \to \tau +1$ ($T$ transformation), this determinant acquires a $-i$ factor, meaning it furnishes the (non-trivial) $\mathbf{\hat{1}}$ representation of $\Gamma_4'$.

\section{Singlet modular forms}
\label{app:forms}

In this appendix we provide a list of all singlet (1D) modular forms available at weights up to 36, for each level $N \leq 5$. The irreps available at each weight $w$ are summarized in~\cref{tab:1forms}. Note that $A_5'$ singlets coincide with the trivial ones available for all $\Gamma_N'$, as no other 1D irreps exist for this group.
These ``trivial singlet'' forms are the well-known $SL(2,\mathbb{Z})$ modular forms generated as polynomials in $E_4$ and $E_6$, with
\begin{equation}
E_{2n}(\tau) = \frac{1}{2\, \zeta(2n)} \sum_{\substack{m,m'\in \mathbb{Z}\\(m,m')\neq (0,0)}}(m+m'\tau)^{-2n} 
\,,
\end{equation}
%
where $n>1$ is an integer.
At a given weight $w>0$, the dimension of the linear space $\mathcal{M}_w(\Gamma)$ of these forms is given by
\begin{equation}
\dim \mathcal{M}_w(\Gamma) =
\begin{cases}
\lfloor{w/12}\rfloor\,,& \!\text{if } w \equiv 2 \!\!\mod{12}\,,\\[1mm]
\lfloor{w/12}\rfloor+1\,,& \!\text{if } w \not\equiv 2 \!\!\mod{12}\,,
\end{cases}
\end{equation}
%
in agreement with the first data column of~\cref{tab:1forms}.

\vskip 2mm

Below we present the first few terms in the $q$-expansions, with $q\equiv \exp(2\pi i \tau)$, of each of the singlet forms $Y_{\mathbf{1}^*}$ whose irreps have been collected in~\cref{tab:1forms}. 
The listed terms are representative: they allow for the unambiguous identification of each form, but are in general insufficient to compute the forms to a good precision in all points of the fundamental domain $\mathcal{D}$.

The fact that $q$-expansion coefficients are real is consistent with the requirement of real superpotential parameters in a \ac{gCP}-invariant modular theory~\cite{Novichkov:2019sqv}.
Each form is normalized according to the global (or integral) normalization based on the Petersson inner product, as advocated in Ref.~\cite{Petcov:2023fwh}. Namely, one has $|\N[Y_{\mathbf{1}^*}]| = 1$ for all these forms, with
\begin{equation} \label{eq:normcusp}
\N\left[Y_{\mathbf{1}^*}^{(w)}(\tau)\right]^2\equiv
\int_\mathcal{D} \left| Y_{\mathbf{1}^*}^{(w)}(x+i y) \right|^2 (2y)^w \frac{dx\, dy}{y^2}
\,,
\end{equation}
%
with $x = \re\tau$ and $y = \im\tau$. \Cref{eq:normcusp} is valid for cusp (singlet) modular forms,
i.e.~forms vanishing in the limit $q\to 0$, in which case the integration is performed over the fundamental domain.
For non-cusp (singlet) forms of weight $w>1$, it should be modified to
\begin{equation} \label{eq:normnoncusp}
\begin{aligned}
\overline{\N}\left[Y_{\mathbf{1}^*}^{(w)}(\tau)\right]^2 \equiv 
\lim_{T\to \infty} \bigg(
\int_{\mathcal{D}_T} \big| Y_{\mathbf{1}^*}^{(w)} \big|^2 (2y)^w \frac{dx\, dy}{y^2} &
\\
 -\, |a_0|^2 \frac{2^w}{w-1} T^{w-1}& \bigg)
\,,
\end{aligned}
\end{equation}
%
where $a_0$ is the constant term in the $q$-expansion of $Y_{\mathbf{1}^*}$
and $\mathcal{D}_T$ is the truncated fundamental domain, with $y < T$ (see also~\cite{Zagier:1981}).

Finally, whenever more than one form is available for a given level, weight and irrep, one is free to choose the basis of the corresponding linear space. Particular basis choices are made below. However, we note that any such choice in general does not commute with the normalization procedure.

\vskip 5mm

\begin{table}[t]
  \centering
  \begin{tabular}{lcccc}
    \toprule
    $w$\quad\,{} & \,\,\,All $\Gamma_N'$\,\,\,{} & \,\, $S_3$ only \,\,{}& $A_4'$ only & \quad \,\,\,$S_4'$ only \quad\,\,\,{} \\
    \midrule
  1 & $-$ 				& $-$ 				& $-$			 	& $-$ \\
  2 & $-$ 				& $-$ 				& $-$ 				& $-$ \\
  3 & $-$ 				& $-$ 				& $-$				& $\mathbf{\hat{1}'}$ \\
  4 & $\mathbf{1}$ 	& $-$				& $\mathbf{1}'$ 		& $-$  \\
  5 & $-$ 				& $-$ 				& $-$ 				& $-$ \\
  6 & $\mathbf{1}$ 	& $\mathbf{1'}$ 		& $-$				& $\mathbf{1'}$		\\
  7 & $-$ 				& $-$ 				& $-$ 				& $\mathbf{\hat{1}'}$ \\
  8 & $\mathbf{1}$ 	& $-$				& $\mathbf{1'}\!,
  					  					      \mathbf{1''}$		& $-$ \\
  9 & $-$				& $-$ 				& $-$ 				& $\mathbf{\hat{1}}, 
										    					      \mathbf{\hat{1}'}$ \\
  10 & $\mathbf{1}$	& $\mathbf{1'}$  		& $\mathbf{1'}$ 		& $\mathbf{1'}$	\\
  11 & $-$ 			& $-$				& $-$ 				& $\mathbf{\hat{1}'}$ 	\\
  12 & $\mathbf{1},
	      \mathbf{1}$ 	& $\mathbf{1'}$		& $\mathbf{1'}\!,
  										      \mathbf{1''}$ 		& $\mathbf{1'}$ \\
  13 & $-$			& $-$				& $-$ 				& $\mathbf{\hat{1}}, 
										  					      \mathbf{\hat{1}'}$		 \\
  14 & $\mathbf{1}$ 	& $\mathbf{1'}$ 		& $\mathbf{1'}\!,
  					  					      \mathbf{1''}$		& $\mathbf{1'}$	\\
  15 & $-$ 			& $-$ 				& $-$ 				& $\mathbf{\hat{1}}, 
															      \mathbf{\hat{1}'}\!, 
															      \mathbf{\hat{1}'}$ \\
  16 & $\mathbf{1},
	      \mathbf{1}$	& $\mathbf{1'}$ 		& $\mathbf{1'}\!,
  						    				      \mathbf{1'}\!,
  					    	   				      \mathbf{1''}$ 		& $\mathbf{1'}$	 \\
  17 & $-$ 			& $-$ 				& $-$ 				& $\mathbf{\hat{1}}, 
															      \mathbf{\hat{1}'}$		\\
  18 & $\mathbf{1},
	     \mathbf{1}$	&	$\mathbf{1'}\!,
	 					  \mathbf{1'}$ 	& $\mathbf{1'}\!,
  					    					      \mathbf{1''}$ 		& $\mathbf{1'}\!, 
										    					      \mathbf{1'}$  \\
  19 & $-$ 			 & $-$  				& $-$ 				& $\mathbf{\hat{1}}, 
															      \mathbf{\hat{1}'}\!, 
										  					      \mathbf{\hat{1}'}$		\\
  20 & $\mathbf{1},
	     \mathbf{1}$	& $\mathbf{1'}$ 		&    $\mathbf{1'}\!,
  										         \mathbf{1'}\!,
  					    					         \mathbf{1''}\!,
  					    					         \mathbf{1''}$ 	& $\mathbf{1'}$  \\
  21 & $-$ 			& $-$ 				& $-$ 				& $\mathbf{\hat{1}}, 
															      \mathbf{\hat{1}}, 
															      \mathbf{\hat{1}'}\!, 
															      \mathbf{\hat{1}'}$ \\
  22 & $\mathbf{1},
	     \mathbf{1}$	 & $\mathbf{1'}\!,
					       \mathbf{1'}$ 		& $\mathbf{1'}\!,
  					    					      \mathbf{1'}\!,
  					  					      \mathbf{1''}$ 		& $\mathbf{1'}\!, 
										    					      \mathbf{1'}$   \\
  23 & $-$ 			& $-$ 				& $-$ 				& $\mathbf{\hat{1}}, 
										   					      \mathbf{\hat{1}'}\!, 
										    					      \mathbf{\hat{1}'}$ \\
  24 & $\mathbf{1},
	     \mathbf{1},
	     \mathbf{1}$	 & $\mathbf{1'}\!,
					       \mathbf{1'}$ 		& $\mathbf{1'}\!,
  					    					      \mathbf{1'}\!,
  					  					      \mathbf{1''}\!,
  					  					      \mathbf{1''}$ 		& $\mathbf{1'}\!, 
										    					      \mathbf{1'}$   \\
  25 & $-$ 			& $-$ 				& $-$ 				& $\mathbf{\hat{1}}, 
										   					      \mathbf{\hat{1}}, 
										   					      \mathbf{\hat{1}'}\!, 
										    					      \mathbf{\hat{1}'}$ \\
  26 & $\mathbf{1},
	     \mathbf{1}$	 & $\mathbf{1'}\!,
					       \mathbf{1'}$ 		& $\mathbf{1'}\!,
  					    					      \mathbf{1'}\!,
  					  					      \mathbf{1''}\!,
  					  					      \mathbf{1''}$ 		& $\mathbf{1'}\!, 
										    					      \mathbf{1'}$   \\
  27 & $-$ 			& $-$ 				& $-$ 				& $\mathbf{\hat{1}}, 
										   					      \mathbf{\hat{1}}, 
										   					      \mathbf{\hat{1}'}\!, 
										    					      \mathbf{\hat{1}'}\!, 
										    					      \mathbf{\hat{1}'}$ \\
  28 & $\mathbf{1},
	     \mathbf{1},
	     \mathbf{1}$	 & $\mathbf{1'}\!,
					       \mathbf{1'}$ 		& $\mathbf{1'}\!,
  					    					      \mathbf{1'}\!,
  					    					      \mathbf{1'}\!,
  					  					      \mathbf{1''}\!,
  					  					      \mathbf{1''}$ 		& $\mathbf{1'}\!, 
										    					      \mathbf{1'}$   \\
  29 & $-$ 			& $-$ 				& $-$ 				& $\mathbf{\hat{1}}, 
										   					      \mathbf{\hat{1}}, 
										   					      \mathbf{\hat{1}'}\!, 
										    					      \mathbf{\hat{1}'}$ \\
  30 & $\mathbf{1},
	     \mathbf{1},
	     \mathbf{1}$	 & $\mathbf{1'}\!,
					       \mathbf{1'}\!,
					       \mathbf{1'}$ 		& $\mathbf{1'}\!,
  					    					      \mathbf{1'}\!,
  					  					      \mathbf{1''}\!,
  					  					      \mathbf{1''}$ 		& $\mathbf{1'}\!, 
										    					      \mathbf{1'}\!, 
										    					      \mathbf{1'}$   \\
  31 & $-$ 			& $-$ 				& $-$ 				& $\mathbf{\hat{1}}, 
										   					      \mathbf{\hat{1}}, 
										   					      \mathbf{\hat{1}'}\!, 
										    					      \mathbf{\hat{1}'}\!, 
										    					      \mathbf{\hat{1}'}$ \\
  32 & $\mathbf{1},
	     \mathbf{1},
	     \mathbf{1}$	 & $\mathbf{1'}\!,
					       \mathbf{1'}$ 		& $\mathbf{1'}\!,
  					    					      \mathbf{1'}\!,
  					    					      \mathbf{1'}\!,
  					  					      \mathbf{1''}\!,
  					  					      \mathbf{1''}\!,
  					  					      \mathbf{1''}$		& $\mathbf{1'}\!, 
										    					      \mathbf{1'}$   \\
  33 & $-$ 			& $-$ 				& $-$ 				& $\mathbf{\hat{1}}, 
										   					      \mathbf{\hat{1}}, 
										    					      \mathbf{\hat{1}}, 
										   					      \mathbf{\hat{1}'}\!, 
										   					      \mathbf{\hat{1}'}\!, 
										    					      \mathbf{\hat{1}'}$ \\
  34 & $\mathbf{1},
	     \mathbf{1},
	     \mathbf{1}$	 & $\mathbf{1'}\!,
					       \mathbf{1'}\!,
					       \mathbf{1'}$ 		& $\mathbf{1'}\!,
  					    					      \mathbf{1'}\!,
  					    					      \mathbf{1'}\!,
  					  					      \mathbf{1''}\!,
  					  					      \mathbf{1''}$ 		& $\mathbf{1'}\!, 
										    					      \mathbf{1'}\!, 
										    					      \mathbf{1'}$   \\
  35 & $-$ 			& $-$ 				& $-$ 				& $\mathbf{\hat{1}}, 
										    					      \mathbf{\hat{1}}, 
										   					      \mathbf{\hat{1}'}\!, 
										   					      \mathbf{\hat{1}'}\!, 
										    					      \mathbf{\hat{1}'}$ \\
  36 & $\mathbf{1},
	     \mathbf{1},
	     \mathbf{1},
	     \mathbf{1}$	 & $\mathbf{1'}\!,
					       \mathbf{1'}\!,
					       \mathbf{1'}$ 		& $\mathbf{1'}\!,
  					    					      \mathbf{1'}\!,
  					    					      \mathbf{1'}\!,
  					  					      \mathbf{1''}\!,
  					  					      \mathbf{1''}\!,
  					  					      \mathbf{1''}$ 		& $\mathbf{1'}\!, 
										    					      \mathbf{1'}\!, 
										    					      \mathbf{1'}$   \\
    \bottomrule
  \end{tabular}
  \caption{Singlet modular form irreps available  at a given weight $w\leq 36$ for the $\Gamma_N'$ groups ($N\leq 5$). 
Hatted (unhatted) irreps are in correspondence with odd (even) weights. 
For weights $w \geq 24$ ($w \geq 36$) one is sure to have at least two (three) independent modular forms furnishing each possible singlet irrep.
}
  \label{tab:1forms}
\end{table}
%

\noindent
\fbox{1D trivial forms, available for all $\Gamma_N'$}
\begingroup \scriptsize
\begin{align*}
Y^{(4)}_{\mathbf{1}} &\simeq 0.489 \left(1 + 240 q + 2160 q^2 + 6720 q^3 + \mathcal{O}(q^{4}) \right) \,,\\
Y^{(6)}_{\mathbf{1}} &\simeq 1.04 \left(1 - 504 q - 16632 q^2 - 122976 q^3 + \mathcal{O}(q^{4}) \right) \,,\\
Y^{(8)}_{\mathbf{1}} &\simeq 0.333 \left(1 + 480 q + 61920 q^2 + 1050240 q^3 + \mathcal{O}(q^{4}) \right) \,,\\
Y^{(10)}_{\mathbf{1}} &\simeq 0.277 \left(1 - 264 q - 135432 q^2 - 5196576 q^3 + \mathcal{O}(q^{4}) \right) \,,\\
Y^{(12)}_{\mathbf{1},1} &\simeq 15.4 \left(q - 24 q^2 + 252 q^3 + \mathcal{O}(q^{4}) \right) \,,\\
Y^{(12)}_{\mathbf{1},2} &\simeq 0.0145 \left(1 - 1008 q + 220752 q^2 + 16519104 q^3 + \mathcal{O}(q^{4}) \right) \,,\\
Y^{(14)}_{\mathbf{1}} &\simeq 0.110 \left(1 - 24 q - 196632 q^2 - 38263776 q^3 + \mathcal{O}(q^{4}) \right) \,,\\
Y^{(16)}_{\mathbf{1},1} &\simeq 2.65 \left(q + 216 q^2 - 3348 q^3 + \mathcal{O}(q^{4}) \right) \,,\\
Y^{(16)}_{\mathbf{1},2} &\simeq 3.49 \times 10^{-3} \left(1 - 768 q - 19008 q^2 + 67329024 q^3 + \mathcal{O}(q^{4}) \right) \,,\\
Y^{(18)}_{\mathbf{1},1} &\simeq 0.911 \left(q - 528 q^2 - 4284 q^3 + \mathcal{O}(q^{4}) \right) \,,\\
Y^{(18)}_{\mathbf{1},2} &\simeq 4.05 \times 10^{-3} \left(1 + 216 q - 200232 q^2 - 85500576 q^3 + \mathcal{O}(q^{4}) \right) \,,\\
Y^{(20)}_{\mathbf{1},1} &\simeq 0.340 \left(q + 456 q^2 + 50652 q^3 + \mathcal{O}(q^{4}) \right) \,,\\
Y^{(20)}_{\mathbf{1},2} &\simeq 6.46 \times 10^{-4} \left(1 - 528 q - 201168 q^2 + 61114944 q^3 + \mathcal{O}(q^{4}) \right) \,,\\
Y^{(22)}_{\mathbf{1},1} &\simeq 0.109 \left(q - 288 q^2 - 128844 q^3 + \mathcal{O}(q^{4}) \right) \,,\\
Y^{(22)}_{\mathbf{1},2} &\simeq 2.38 \times 10^{-4} \left(1 + 456 q - 146232 q^2 - 133082976 q^3 + \mathcal{O}(q^{4}) \right) \,,\\
Y^{(24)}_{\mathbf{1},1} &\simeq 145 \left(q^2 - 48 q^3 + \mathcal{O}(q^{4}) \right) \,,\\
Y^{(24)}_{\mathbf{1},2} &\simeq 0.0316 \left(q + 696 q^2 + 162252 q^3 + \mathcal{O}(q^{4}) \right) \,,\\
Y^{(24)}_{\mathbf{1},3} &\simeq 1.08 \times 10^{-4} \left(1 - 288 q - 325728 q^2 + 11700864 q^3 + \mathcal{O}(q^{4}) \right) \,,\\
Y^{(26)}_{\mathbf{1},1} &\simeq 8.51 \times 10^{-3} \left(q - 48 q^2 - 195804 q^3 + \mathcal{O}(q^{4}) \right) \,,\\
Y^{(26)}_{\mathbf{1},2} &\simeq 1.22 \times 10^{-5} \left(1 + 696 q - 34632 q^2 - 167186976 q^3 + \mathcal{O}(q^{4}) \right) \,,\\
Y^{(28)}_{\mathbf{1},1} &\simeq 29.3 \left(q^2 + 192 q^3 + \mathcal{O}(q^{4}) \right) \,,\\
Y^{(28)}_{\mathbf{1},2} &\simeq 2.09 \times 10^{-3} \left(q + 936 q^2 + 331452 q^3 + \mathcal{O}(q^{4}) \right) \,,\\
Y^{(28)}_{\mathbf{1},3} &\simeq 3.77 \times 10^{-5} \left(1 - 48 q - 392688 q^2 - 67089216 q^3 + \mathcal{O}(q^{4}) \right) \,,\\
Y^{(30)}_{\mathbf{1},1} &\simeq 10.2 \left(q^2 - 552 q^3 + \mathcal{O}(q^{4}) \right) \,,\\
Y^{(30)}_{\mathbf{1},2} &\simeq 4.79 \times 10^{-4} \left(q + 192 q^2 - 205164 q^3 + \mathcal{O}(q^{4}) \right) \,,\\
Y^{(30)}_{\mathbf{1},3} &\simeq 5.12 \times 10^{-7} \left(1 + 936 q + 134568 q^2 - 173988576 q^3 + \mathcal{O}(q^{4}) \right) \,,\\
Y^{(32)}_{\mathbf{1},1} &\simeq 4.83 \left(q^2 + 432 q^3 + \mathcal{O}(q^{4}) \right) \,,\\
Y^{(32)}_{\mathbf{1},2} &\simeq 1.02 \times 10^{-4} \left(q + 1176 q^2 + 558252 q^3 + \mathcal{O}(q^{4}) \right) \,,\\
Y^{(32)}_{\mathbf{1},3} &\simeq 5.31 \times 10^{-7} \left(1 + 192 q - 402048 q^2 - 161431296 q^3 + \mathcal{O}(q^{4}) \right) \,,\\
Y^{(34)}_{\mathbf{1},1} &\simeq 1.88 \left(q^2 - 312 q^3 + \mathcal{O}(q^{4}) \right) \,,\\
Y^{(34)}_{\mathbf{1},2} &\simeq 2.03 \times 10^{-5} \left(q + 432 q^2 - 156924 q^3 + \mathcal{O}(q^{4}) \right) \,,\\
Y^{(34)}_{\mathbf{1},3} &\simeq 1.73 \times 10^{-8} \left(1 + 1176 q + 361368 q^2 - 139663776 q^3 + \mathcal{O}(q^{4}) \right) \,,\\
Y^{(36)}_{\mathbf{1},1} &\simeq 1290 \left(q^3 + \mathcal{O}(q^{4}) \right) \,,\\
Y^{(36)}_{\mathbf{1},2} &\simeq 0.683 \left(q^2 + 672 q^3 + \mathcal{O}(q^{4}) \right) \,,\\
Y^{(36)}_{\mathbf{1},3} &\simeq 3.82 \times 10^{-6} \left(q + 1416 q^2 + 842652 q^3 + \mathcal{O}(q^{4}) \right) \,,\\
Y^{(36)}_{\mathbf{1},4} &\simeq 8.85 \times 10^{-9} \left(1 + 432 q - 353808 q^2 - 257501376 q^3 + \mathcal{O}(q^{4}) \right) \,.
\end{align*}
\endgroup
%

\noindent
\fbox{1D forms for $\Gamma_2 \simeq S_3$}
\begingroup \scriptsize
\begin{align*}
Y^{(6)}_{\mathbf{1'}} &\simeq 4.72 \left(\sqrt{q} - 12 q^{3/2} + 54 q^{5/2} + \mathcal{O}(q^{7/2}) \right) \,,\\
Y^{(10)}_{\mathbf{1'}} &\simeq 0.661 \left(\sqrt{q} + 228 q^{3/2} - 666 q^{5/2} + \mathcal{O}(q^{7/2}) \right) \,,\\
Y^{(12)}_{\mathbf{1'}} &\simeq 0.199 \left(\sqrt{q} - 516 q^{3/2} - 10530 q^{5/2} + \mathcal{O}(q^{7/2}) \right) \,,\\
Y^{(14)}_{\mathbf{1'}} &\simeq 0.0551 \left(\sqrt{q} + 468 q^{3/2} + 56214 q^{5/2} + \mathcal{O}(q^{7/2}) \right) \,,\\
Y^{(16)}_{\mathbf{1'}} &\simeq 0.0128 \left(\sqrt{q} - 276 q^{3/2} - 132210 q^{5/2} + \mathcal{O}(q^{7/2}) \right) \,,\\
Y^{(18)}_{\mathbf{1'},1} &\simeq 2.59 \times 10^{-3} \left(\sqrt{q} + 708 q^{3/2} + 170694 q^{5/2} + \mathcal{O}(q^{7/2}) \right) \,,\\
Y^{(18)}_{\mathbf{1'},2} &\simeq 2.59 \times 10^{-3} \left(\sqrt{q} - 1020 q^{3/2} + 232902 q^{5/2} + \mathcal{O}(q^{7/2}) \right) \,,\\
Y^{(20)}_{\mathbf{1'}} &\simeq 4.66 \times 10^{-4} \left(\sqrt{q} - 36 q^{3/2} - 196290 q^{5/2} + \mathcal{O}(q^{7/2}) \right) \,,\\
Y^{(22)}_{\mathbf{1'},1} &\simeq 7.52 \times 10^{-5} \left(\sqrt{q} + 948 q^{3/2} + 342774 q^{5/2} + \mathcal{O}(q^{7/2}) \right) \,,\\
Y^{(22)}_{\mathbf{1'},2} &\simeq 7.52 \times 10^{-5} \left(\sqrt{q} - 780 q^{3/2} - 9738 q^{5/2} + \mathcal{O}(q^{7/2}) \right) \,,\\
Y^{(24)}_{\mathbf{1'},1} &\simeq 1.09 \times 10^{-5} \left(\sqrt{q} - 1524 q^{3/2} + 730350 q^{5/2} + \mathcal{O}(q^{7/2}) \right) \,,\\
Y^{(24)}_{\mathbf{1'},2} &\simeq 1.09 \times 10^{-5} \left(\sqrt{q} + 204 q^{3/2} - 202770 q^{5/2} + \mathcal{O}(q^{7/2}) \right) \,,\\
Y^{(26)}_{\mathbf{1'},1} &\simeq 1.47 \times 10^{-6} \left(\sqrt{q} + 1188 q^{3/2} + 572454 q^{5/2} + \mathcal{O}(q^{7/2}) \right) \,,\\
Y^{(26)}_{\mathbf{1'},2} &\simeq 1.47 \times 10^{-6} \left(\sqrt{q} - 540 q^{3/2} - 194778 q^{5/2} + \mathcal{O}(q^{7/2}) \right) \,,\\
Y^{(28)}_{\mathbf{1'},1} &\simeq 1.81 \times 10^{-7} \left(\sqrt{q} - 1284 q^{3/2} + 366750 q^{5/2} + \mathcal{O}(q^{7/2}) \right) \,,\\
Y^{(28)}_{\mathbf{1'},2} &\simeq 1.81 \times 10^{-7} \left(\sqrt{q} + 444 q^{3/2} - 151650 q^{5/2} + \mathcal{O}(q^{7/2}) \right) \,,\\
Y^{(30)}_{\mathbf{1'},1} &\simeq 2.07 \times 10^{-8} \left(\sqrt{q} + 1428 q^{3/2} + 859734 q^{5/2} + \mathcal{O}(q^{7/2}) \right) \,,\\
Y^{(30)}_{\mathbf{1'},2} &\simeq 2.07 \times 10^{-8} \left(\sqrt{q} - 2028 q^{3/2} + 1481814 q^{5/2} + \mathcal{O}(q^{7/2}) \right) \,,\\
Y^{(30)}_{\mathbf{1'},3} &\simeq 2.07 \times 10^{-8} \left(\sqrt{q} - 300 q^{3/2} - 322218 q^{5/2} + \mathcal{O}(q^{7/2}) \right) \,,\\
Y^{(32)}_{\mathbf{1'},1} &\simeq 2.20 \times 10^{-9} \left(\sqrt{q} - 1044 q^{3/2} + 60750 q^{5/2} + \mathcal{O}(q^{7/2}) \right) \,,\\
Y^{(32)}_{\mathbf{1'},2} &\simeq 2.20 \times 10^{-9} \left(\sqrt{q} + 684 q^{3/2} - 42930 q^{5/2} + \mathcal{O}(q^{7/2}) \right) \,,\\
Y^{(34)}_{\mathbf{1'},1} &\simeq 2.19 \times 10^{-10} \left(\sqrt{q} + 1668 q^{3/2} + 1204614 q^{5/2} + \mathcal{O}(q^{7/2}) \right) \,,\\
Y^{(34)}_{\mathbf{1'},2} &\simeq 2.19 \times 10^{-10} \left(\sqrt{q} - 60 q^{3/2} - 392058 q^{5/2} + \mathcal{O}(q^{7/2}) \right) \,,\\
Y^{(34)}_{\mathbf{1'},3} &\simeq 2.19 \times 10^{-10} \left(\sqrt{q} - 1788 q^{3/2} + 997254 q^{5/2} + \mathcal{O}(q^{7/2}) \right) \,,\\
Y^{(36)}_{\mathbf{1'},1} &\simeq 2.06 \times 10^{-11} \left(\sqrt{q} - 804 q^{3/2} - 187650 q^{5/2} + \mathcal{O}(q^{7/2}) \right) \,,\\
Y^{(36)}_{\mathbf{1'},2} &\simeq 2.06 \times 10^{-11} \left(\sqrt{q} - 2532 q^{3/2} + 2487294 q^{5/2} + \mathcal{O}(q^{7/2}) \right) \,,\\
Y^{(36)}_{\mathbf{1'},3} &\simeq 2.06 \times 10^{-11} \left(\sqrt{q} + 924 q^{3/2} + 123390 q^{5/2} + \mathcal{O}(q^{7/2}) \right) \,.
\end{align*}
\endgroup
%

\noindent
\fbox{1D forms for $\Gamma_3' \simeq A_4'$}
\begingroup \scriptsize
\begin{align*}
Y^{(4)}_{\mathbf{1'}} &\simeq 3.10 \left(q^{1/3} - 8 q^{4/3} + 20 q^{7/3} + \mathcal{O}(q^{10/3}) \right) \,,\\
Y^{(8)}_{\mathbf{1'}} &\simeq 0.372 \left(q^{1/3} + 232 q^{4/3} + 260 q^{7/3} + \mathcal{O}(q^{10/3}) \right) \,,\\
Y^{(8)}_{\mathbf{1''}} &\simeq 7.05 \left(q^{2/3} - 16 q^{5/3} + 104 q^{8/3} + \mathcal{O}(q^{10/3}) \right) \,,\\
Y^{(10)}_{\mathbf{1'}} &\simeq 0.0976 \left(q^{1/3} - 512 q^{4/3} - 12580 q^{7/3} + \mathcal{O}(q^{10/3}) \right) \,,\\
Y^{(12)}_{\mathbf{1'}} &\simeq 0.0217 \left(q^{1/3} + 472 q^{4/3} + 58100 q^{7/3} + \mathcal{O}(q^{10/3}) \right) \,,\\
Y^{(12)}_{\mathbf{1''}} &\simeq 1.09 \left(q^{2/3} + 224 q^{5/3} - 1576 q^{8/3} + \mathcal{O}(q^{10/3}) \right) \,,\\
Y^{(14)}_{\mathbf{1'}} &\simeq 3.94 \times 10^{-3} \left(q^{1/3} - 272 q^{4/3} - 133300 q^{7/3} + \mathcal{O}(q^{10/3}) \right) \,,\\
Y^{(14)}_{\mathbf{1''}} &\simeq 0.351 \left(q^{2/3} - 520 q^{5/3} - 8464 q^{8/3} + \mathcal{O}(q^{10/3}) \right) \,,\\
Y^{(16)}_{\mathbf{1'},1} &\simeq 32.8 \left(q^{4/3} - 32 q^{7/3} + \mathcal{O}(q^{10/3}) \right) \,,\\
Y^{(16)}_{\mathbf{1'},2} &\simeq 6.12 \times 10^{-4} \left(q^{1/3} + 712 q^{4/3} + 173540 q^{7/3} + \mathcal{O}(q^{10/3}) \right) \,,\\
Y^{(16)}_{\mathbf{1''}} &\simeq 0.111 \left(q^{2/3} + 464 q^{5/3} + 54344 q^{8/3} + \mathcal{O}(q^{10/3}) \right) \,,\\
Y^{(18)}_{\mathbf{1'}} &\simeq 8.28 \times 10^{-5} \left(q^{1/3} - 32 q^{4/3} - 196420 q^{7/3} + \mathcal{O}(q^{10/3}) \right) \,,\\
Y^{(18)}_{\mathbf{1''}} &\simeq 0.0300 \left(q^{2/3} - 280 q^{5/3} - 131104 q^{8/3} + \mathcal{O}(q^{10/3}) \right) \,,\\
Y^{(20)}_{\mathbf{1'},1} &\simeq 6.07 \left(q^{4/3} + 208 q^{7/3} + \mathcal{O}(q^{10/3}) \right) \,,\\
Y^{(20)}_{\mathbf{1'},2} &\simeq 9.91 \times 10^{-6} \left(q^{1/3} + 952 q^{4/3} + 346580 q^{7/3} + \mathcal{O}(q^{10/3}) \right) \,,\\
Y^{(20)}_{\mathbf{1''},1} &\simeq 69.1 \left(q^{5/3} - 40 q^{8/3} + \mathcal{O}(q^{10/3}) \right) \,,\\
Y^{(20)}_{\mathbf{1''},2} &\simeq 7.16 \times 10^{-3} \left(q^{2/3} + 704 q^{5/3} + 167864 q^{8/3} + \mathcal{O}(q^{10/3}) \right) \,,\\
Y^{(22)}_{\mathbf{1'},1} &\simeq 2.13 \left(q^{4/3} - 536 q^{7/3} + \mathcal{O}(q^{10/3}) \right) \,,\\
Y^{(22)}_{\mathbf{1'},2} &\simeq 1.06 \times 10^{-6} \left(q^{1/3} + 208 q^{4/3} - 201940 q^{7/3} + \mathcal{O}(q^{10/3}) \right) \,,\\
Y^{(22)}_{\mathbf{1''}} &\simeq 1.54 \times 10^{-3} \left(q^{2/3} - 40 q^{5/3} - 196144 q^{8/3} + \mathcal{O}(q^{10/3}) \right) \,,\\
Y^{(24)}_{\mathbf{1'},1} &\simeq 0.879 \left(q^{4/3} + 448 q^{7/3} + \mathcal{O}(q^{10/3}) \right) \,,\\
Y^{(24)}_{\mathbf{1'},2} &\simeq 1.03 \times 10^{-7} \left(q^{1/3} + 1192 q^{4/3} + 577220 q^{7/3} + \mathcal{O}(q^{10/3}) \right) \,,\\
Y^{(24)}_{\mathbf{1''},1} &\simeq 13.5 \left(q^{5/3} + 200 q^{8/3} + \mathcal{O}(q^{10/3}) \right) \,,\\
Y^{(24)}_{\mathbf{1''},2} &\simeq 3.00 \times 10^{-4} \left(q^{2/3} + 944 q^{5/3} + 338984 q^{8/3} + \mathcal{O}(q^{10/3}) \right) \,,\\
Y^{(26)}_{\mathbf{1'},1} &\simeq 0.310 \left(q^{4/3} - 296 q^{7/3} + \mathcal{O}(q^{10/3}) \right) \,,\\
Y^{(26)}_{\mathbf{1'},2} &\simeq 9.25 \times 10^{-9} \left(q^{1/3} + 448 q^{4/3} - 149860 q^{7/3} + \mathcal{O}(q^{10/3}) \right) \,,\\
Y^{(26)}_{\mathbf{1''},1} &\simeq 4.73 \left(q^{5/3} - 544 q^{8/3} + \mathcal{O}(q^{10/3}) \right) \,,\\
Y^{(26)}_{\mathbf{1''},2} &\simeq 5.36 \times 10^{-5} \left(q^{2/3} + 200 q^{5/3} - 203584 q^{8/3} + \mathcal{O}(q^{10/3}) \right) \,,\\
Y^{(28)}_{\mathbf{1'},1} &\simeq 301 \left(q^{7/3} + \mathcal{O}(q^{10/3}) \right) \,,\\
Y^{(28)}_{\mathbf{1'},2} &\simeq 0.100 \left(q^{4/3} + 688 q^{7/3} + \mathcal{O}(q^{10/3}) \right) \,,\\
Y^{(28)}_{\mathbf{1'},3} &\simeq 7.60 \times 10^{-10} \left(q^{1/3} + 1432 q^{4/3} + 865460 q^{7/3} + \mathcal{O}(q^{10/3}) \right) \,,\\
Y^{(28)}_{\mathbf{1''},1} &\simeq 2.10 \left(q^{5/3} + 440 q^{8/3} + \mathcal{O}(q^{10/3}) \right) \,,\\
Y^{(28)}_{\mathbf{1''},2} &\simeq 8.80 \times 10^{-6} \left(q^{2/3} + 1184 q^{5/3} + 567704 q^{8/3} + \mathcal{O}(q^{10/3}) \right) \,,\\
Y^{(30)}_{\mathbf{1'},1} &\simeq 0.0311 \left(q^{4/3} - 56 q^{7/3} + \mathcal{O}(q^{10/3}) \right) \,,\\
Y^{(30)}_{\mathbf{1'},2} &\simeq 5.79 \times 10^{-11} \left(q^{1/3} + 688 q^{4/3} - 40180 q^{7/3} + \mathcal{O}(q^{10/3}) \right) \,,\\
Y^{(30)}_{\mathbf{1''},1} &\simeq 0.788 \left(q^{5/3} - 304 q^{8/3} + \mathcal{O}(q^{10/3}) \right) \,,\\
Y^{(30)}_{\mathbf{1''},2} &\simeq 1.34 \times 10^{-6} \left(q^{2/3} + 440 q^{5/3} - 153424 q^{8/3} + \mathcal{O}(q^{10/3}) \right) \,,\\
Y^{(32)}_{\mathbf{1'},1} &\simeq 63.0 \left(q^{7/3} + \mathcal{O}(q^{10/3}) \right) \,,\\
Y^{(32)}_{\mathbf{1'},2} &\simeq 8.75 \times 10^{-3} \left(q^{4/3} + 928 q^{7/3} + \mathcal{O}(q^{10/3}) \right) \,,\\
Y^{(32)}_{\mathbf{1'},3} &\simeq 4.11 \times 10^{-12} \left(q^{1/3} - 56 q^{4/3} - 392284 q^{7/3} + \mathcal{O}(q^{10/3}) \right) \,,\\
Y^{(32)}_{\mathbf{1''},1} &\simeq 623 \left(q^{8/3} + \mathcal{O}(q^{10/3}) \right) \,,\\
Y^{(32)}_{\mathbf{1''},2} &\simeq 0.273 \left(q^{5/3} + 680 q^{8/3} + \mathcal{O}(q^{10/3}) \right) \,,\\
Y^{(32)}_{\mathbf{1''},3} &\simeq 1.90 \times 10^{-7} \left(q^{2/3} + 1424 q^{5/3} + 854024 q^{8/3} + \mathcal{O}(q^{10/3}) \right) \,,\\
Y^{(34)}_{\mathbf{1'},1} &\simeq 21.7 \left(q^{7/3} + \mathcal{O}(q^{10/3}) \right) \,,\\
Y^{(34)}_{\mathbf{1'},2} &\simeq 2.34 \times 10^{-3} \left(q^{4/3} + 184 q^{7/3} + \mathcal{O}(q^{10/3}) \right) \,,\\
Y^{(34)}_{\mathbf{1'},3} &\simeq 2.73 \times 10^{-13} \left(q^{1/3} + 928 q^{4/3} + 127100 q^{7/3} + \mathcal{O}(q^{10/3}) \right) \,,\\
Y^{(34)}_{\mathbf{1''},1} &\simeq 0.0933 \left(q^{5/3} - 64 q^{8/3} + \mathcal{O}(q^{10/3}) \right) \,,\\
Y^{(34)}_{\mathbf{1''},2} &\simeq 2.53 \times 10^{-8} \left(q^{2/3} + 680 q^{5/3} - 45664 q^{8/3} + \mathcal{O}(q^{10/3}) \right) \,,\\
Y^{(36)}_{\mathbf{1'},1} &\simeq 10.8 \left(q^{7/3} + \mathcal{O}(q^{10/3}) \right) \,,\\
Y^{(36)}_{\mathbf{1'},2} &\simeq 5.86 \times 10^{-4} \left(q^{4/3} + 1168 q^{7/3} + \mathcal{O}(q^{10/3}) \right) \,,\\
Y^{(36)}_{\mathbf{1'},3} &\simeq 1.70 \times 10^{-14} \left(q^{1/3} + 184 q^{4/3} - 403564 q^{7/3} + \mathcal{O}(q^{10/3}) \right) \,,\\
Y^{(36)}_{\mathbf{1''},1} &\simeq 134 \left(q^{8/3} + \mathcal{O}(q^{10/3}) \right) \,,\\
Y^{(36)}_{\mathbf{1''},2} &\simeq 0.0287 \left(q^{5/3} - 808 q^{8/3} + \mathcal{O}(q^{10/3}) \right) \,,\\
Y^{(36)}_{\mathbf{1''},3} &\simeq 3.16 \times 10^{-9} \left(q^{2/3} + 1664 q^{5/3} + 1197944 q^{8/3} + \mathcal{O}(q^{10/3}) \right) \,.
\end{align*}
\endgroup
%

\noindent
\fbox{1D forms for $\Gamma_4' \simeq S_4'$}
\begingroup \scriptsize
\begin{align*}
Y^{(3)}_{\mathbf{\hat{1}'}} &\simeq 2.49 \left(q^{1/4} - 6 q^{5/4} + 9 q^{9/4} + \mathcal{O}(q^{13/4}) \right) \,,\\
Y^{(6)}_{\mathbf{1'}} &\simeq 4.72 \left(\sqrt{q} - 12 q^{3/2} + 54 q^{5/2} + \mathcal{O}(q^{7/2}) \right) \,,\\
Y^{(7)}_{\mathbf{\hat{1}'}} &\simeq 0.261 \left(q^{1/4} + 234 q^{5/4} + 729 q^{9/4} + \mathcal{O}(q^{13/4}) \right) \,,\\
Y^{(9)}_{\mathbf{\hat{1}}} &\simeq 8.59 \left(q^{3/4} - 18 q^{7/4} + 135 q^{11/4} + \mathcal{O}(q^{13/4}) \right) \,,\\
Y^{(9)}_{\mathbf{\hat{1}'}} &\simeq 0.0605 \left(q^{1/4} - 510 q^{5/4} - 13599 q^{9/4} + \mathcal{O}(q^{13/4}) \right) \,,\\
Y^{(10)}_{\mathbf{1'}} &\simeq 0.661 \left(\sqrt{q} + 228 q^{3/2} - 666 q^{5/2} + \mathcal{O}(q^{7/2}) \right) \,,\\
Y^{(11)}_{\mathbf{\hat{1}'}} &\simeq 0.0112 \left(q^{1/4} + 474 q^{5/4} + 59049 q^{9/4} + \mathcal{O}(q^{13/4}) \right) \,,\\
Y^{(12)}_{\mathbf{1'}} &\simeq 0.199 \left(\sqrt{q} - 516 q^{3/2} - 10530 q^{5/2} + \mathcal{O}(q^{7/2}) \right) \,,\\
Y^{(13)}_{\mathbf{\hat{1}}} &\simeq 1.37 \left(q^{3/4} + 222 q^{7/4} - 2025 q^{11/4} + \mathcal{O}(q^{13/4}) \right) \,,\\
Y^{(13)}_{\mathbf{\hat{1}'}} &\simeq 1.68 \times 10^{-3} \left(q^{1/4} - 270 q^{5/4} - 133839 q^{9/4} + \mathcal{O}(q^{13/4}) \right) \,,\\
Y^{(14)}_{\mathbf{1'}} &\simeq 0.0551 \left(\sqrt{q} + 468 q^{3/2} + 56214 q^{5/2} + \mathcal{O}(q^{7/2}) \right) \,,\\
Y^{(15)}_{\mathbf{\hat{1}}} &\simeq 0.453 \left(q^{3/4} - 522 q^{7/4} - 7425 q^{11/4} + \mathcal{O}(q^{13/4}) \right) \,,\\
Y^{(15)}_{\mathbf{\hat{1}'},1} &\simeq 27.1 \left(q^{5/4} - 30 q^{9/4} + \mathcal{O}(q^{13/4}) \right) \,,\\
Y^{(15)}_{\mathbf{\hat{1}'},2} &\simeq 2.11 \times 10^{-4} \left(q^{1/4} - 1014 q^{5/4} + 226809 q^{9/4} + \mathcal{O}(q^{13/4}) \right) \,,\\
Y^{(16)}_{\mathbf{1'}} &\simeq 0.0128 \left(\sqrt{q} - 276 q^{3/2} - 132210 q^{5/2} + \mathcal{O}(q^{7/2}) \right) \,,\\
Y^{(17)}_{\mathbf{\hat{1}}} &\simeq 0.151 \left(q^{3/4} + 462 q^{7/4} + 53415 q^{11/4} + \mathcal{O}(q^{13/4}) \right) \,,\\
Y^{(17)}_{\mathbf{\hat{1}'}} &\simeq 2.29 \times 10^{-5} \left(q^{1/4} - 30 q^{5/4} - 196479 q^{9/4} + \mathcal{O}(q^{13/4}) \right) \,,\\
Y^{(18)}_{\mathbf{1'},1} &\simeq 47.6 \left(q^{3/2} - 36 q^{5/2} + \mathcal{O}(q^{7/2}) \right) \,,\\
Y^{(18)}_{\mathbf{1'},2} &\simeq 2.59 \times 10^{-3} \left(\sqrt{q} - 1020 q^{3/2} + 232902 q^{5/2} + \mathcal{O}(q^{7/2}) \right) \,,\\
Y^{(19)}_{\mathbf{\hat{1}}} &\simeq 0.0430 \left(q^{3/4} - 282 q^{7/4} - 130545 q^{11/4} + \mathcal{O}(q^{13/4}) \right) \,,\\
Y^{(19)}_{\mathbf{\hat{1}'},1} &\simeq 4.96 \left(q^{5/4} + 210 q^{9/4} + \mathcal{O}(q^{13/4}) \right) \,,\\
Y^{(19)}_{\mathbf{\hat{1}'},2} &\simeq 2.18 \times 10^{-6} \left(q^{1/4} - 774 q^{5/4} - 14391 q^{9/4} + \mathcal{O}(q^{13/4}) \right) \,,\\
Y^{(20)}_{\mathbf{1'}} &\simeq 4.66 \times 10^{-4} \left(\sqrt{q} - 36 q^{3/2} - 196290 q^{5/2} + \mathcal{O}(q^{7/2}) \right) \,,\\
Y^{(21)}_{\mathbf{\hat{1}},1} &\simeq 0.0109 \left(q^{3/4} + 702 q^{7/4} + 166455 q^{11/4} + \mathcal{O}(q^{13/4}) \right) \,,\\
Y^{(21)}_{\mathbf{\hat{1}},2} &\simeq 0.0108 \left(q^{3/4} - 1026 q^{7/4} + 239031 q^{11/4} + \mathcal{O}(q^{13/4}) \right) \,,\\
Y^{(21)}_{\mathbf{\hat{1}'},1} &\simeq 1.73 \left(q^{5/4} - 534 q^{9/4} + \mathcal{O}(q^{13/4}) \right) \,,\\
Y^{(21)}_{\mathbf{\hat{1}'},2} &\simeq 1.85 \times 10^{-7} \left(q^{1/4} + 210 q^{5/4} - 201519 q^{9/4} + \mathcal{O}(q^{13/4}) \right) \,,\\
Y^{(22)}_{\mathbf{1'},1} &\simeq 9.07 \left(q^{3/2} + 204 q^{5/2} + \mathcal{O}(q^{7/2}) \right) \,,\\
Y^{(22)}_{\mathbf{1'},2} &\simeq 7.52 \times 10^{-5} \left(\sqrt{q} - 780 q^{3/2} - 9738 q^{5/2} + \mathcal{O}(q^{7/2}) \right) \,,\\
Y^{(23)}_{\mathbf{\hat{1}}} &\simeq 2.51 \times 10^{-3} \left(q^{3/4} - 42 q^{7/4} - 196065 q^{11/4} + \mathcal{O}(q^{13/4}) \right) \,,\\
Y^{(23)}_{\mathbf{\hat{1}'},1} &\simeq 0.699 \left(q^{5/4} + 450 q^{9/4} + \mathcal{O}(q^{13/4}) \right) \,,\\
Y^{(23)}_{\mathbf{\hat{1}'},2} &\simeq 1.42 \times 10^{-8} \left(q^{1/4} - 534 q^{5/4} - 197991 q^{9/4} + \mathcal{O}(q^{13/4}) \right) \,,\\
Y^{(24)}_{\mathbf{1'},1} &\simeq 3.19 \left(q^{3/2} - 540 q^{5/2} + \mathcal{O}(q^{7/2}) \right) \,,\\
Y^{(24)}_{\mathbf{1'},2} &\simeq 1.09 \times 10^{-5} \left(\sqrt{q} + 204 q^{3/2} - 202770 q^{5/2} + \mathcal{O}(q^{7/2}) \right) \,,\\
Y^{(25)}_{\mathbf{\hat{1}},1} &\simeq 16.4 \left(q^{7/4} + 198 q^{11/4} + \mathcal{O}(q^{13/4}) \right) \,,\\
Y^{(25)}_{\mathbf{\hat{1}},2} &\simeq 5.27 \times 10^{-4} \left(q^{3/4} + 942 q^{7/4} + 337095 q^{11/4} + \mathcal{O}(q^{13/4}) \right) \,,\\
Y^{(25)}_{\mathbf{\hat{1}'},1} &\simeq 0.242 \left(q^{5/4} - 294 q^{9/4} + \mathcal{O}(q^{13/4}) \right) \,,\\
Y^{(25)}_{\mathbf{\hat{1}'},2} &\simeq 9.92 \times 10^{-10} \left(q^{1/4} + 450 q^{5/4} - 148959 q^{9/4} + \mathcal{O}(q^{13/4}) \right) \,,\\
Y^{(26)}_{\mathbf{1'},1} &\simeq 1.37 \left(q^{3/2} + 444 q^{5/2} + \mathcal{O}(q^{7/2}) \right) \,,\\
Y^{(26)}_{\mathbf{1'},2} &\simeq 1.47 \times 10^{-6} \left(\sqrt{q} - 540 q^{3/2} - 194778 q^{5/2} + \mathcal{O}(q^{7/2}) \right) \,,\\
Y^{(27)}_{\mathbf{\hat{1}},1} &\simeq 5.75 \left(q^{7/4} - 546 q^{11/4} + \mathcal{O}(q^{13/4}) \right) \,,\\
Y^{(27)}_{\mathbf{\hat{1}},2} &\simeq 1.01 \times 10^{-4} \left(q^{3/4} + 198 q^{7/4} - 203985 q^{11/4} + \mathcal{O}(q^{13/4}) \right) \,,\\
Y^{(27)}_{\mathbf{\hat{1}'},1} &\simeq 251 \left(q^{9/4} + \mathcal{O}(q^{13/4}) \right) \,,\\
Y^{(27)}_{\mathbf{\hat{1}'},2} &\simeq 0.0766 \left(q^{5/4} + 690 q^{9/4} + \mathcal{O}(q^{13/4}) \right) \,,\\
Y^{(27)}_{\mathbf{\hat{1}'},3} &\simeq 6.36 \times 10^{-11} \left(q^{1/4} - 294 q^{5/4} - 323991 q^{9/4} + \mathcal{O}(q^{13/4}) \right) \,,\\
Y^{(28)}_{\mathbf{1'},1} &\simeq 0.499 \left(q^{3/2} - 300 q^{5/2} + \mathcal{O}(q^{7/2}) \right) \,,\\
Y^{(28)}_{\mathbf{1'},2} &\simeq 1.81 \times 10^{-7} \left(\sqrt{q} + 444 q^{3/2} - 151650 q^{5/2} + \mathcal{O}(q^{7/2}) \right) \,,\\
Y^{(29)}_{\mathbf{\hat{1}},1} &\simeq 2.60 \left(q^{7/4} + 438 q^{11/4} + \mathcal{O}(q^{13/4}) \right) \,,\\
Y^{(29)}_{\mathbf{\hat{1}},2} &\simeq 1.80 \times 10^{-5} \left(q^{3/4} - 546 q^{7/4} - 191529 q^{11/4} + \mathcal{O}(q^{13/4}) \right) \,,\\
Y^{(29)}_{\mathbf{\hat{1}'},1} &\simeq 0.0230 \left(q^{5/4} - 54 q^{9/4} + \mathcal{O}(q^{13/4}) \right) \,,\\
Y^{(29)}_{\mathbf{\hat{1}'},2} &\simeq 3.77 \times 10^{-12} \left(q^{1/4} + 690 q^{5/4} - 38799 q^{9/4} + \mathcal{O}(q^{13/4}) \right) \,,\\
Y^{(30)}_{\mathbf{1'},1} &\simeq 433 \left(q^{5/2} + \mathcal{O}(q^{7/2}) \right) \,,\\
Y^{(30)}_{\mathbf{1'},2} &\simeq 0.168 \left(q^{3/2} + 684 q^{5/2} + \mathcal{O}(q^{7/2}) \right) \,,\\
Y^{(30)}_{\mathbf{1'},3} &\simeq 2.07 \times 10^{-8} \left(\sqrt{q} - 300 q^{3/2} - 322218 q^{5/2} + \mathcal{O}(q^{7/2}) \right) \,,\\
Y^{(31)}_{\mathbf{\hat{1}},1} &\simeq 0.984 \left(q^{7/4} - 306 q^{11/4} + \mathcal{O}(q^{13/4}) \right) \,,\\
Y^{(31)}_{\mathbf{\hat{1}},2} &\simeq 2.98 \times 10^{-6} \left(q^{3/4} + 438 q^{7/4} - 154305 q^{11/4} + \mathcal{O}(q^{13/4}) \right) \,,\\
Y^{(31)}_{\mathbf{\hat{1}'},1} &\simeq 52.1 \left(q^{9/4} + \mathcal{O}(q^{13/4}) \right) \,,\\
Y^{(31)}_{\mathbf{\hat{1}'},2} &\simeq 6.31 \times 10^{-3} \left(q^{5/4} - 798 q^{9/4} + \mathcal{O}(q^{13/4}) \right) \,,\\
Y^{(31)}_{\mathbf{\hat{1}'},3} &\simeq 2.07 \times 10^{-13} \left(q^{1/4} - 54 q^{5/4} - 392391 q^{9/4} + \mathcal{O}(q^{13/4}) \right) \,,\\
Y^{(32)}_{\mathbf{1'},1} &\simeq 0.0548 \left(q^{3/2} - 60 q^{5/2} + \mathcal{O}(q^{7/2}) \right) \,,\\
Y^{(32)}_{\mathbf{1'},2} &\simeq 2.20 \times 10^{-9} \left(\sqrt{q} + 684 q^{3/2} - 42930 q^{5/2} + \mathcal{O}(q^{7/2}) \right) \,,\\
Y^{(33)}_{\mathbf{\hat{1}},1} &\simeq 747 \left(q^{11/4} + \mathcal{O}(q^{13/4}) \right) \,,\\
Y^{(33)}_{\mathbf{\hat{1}},2} &\simeq 0.346 \left(q^{7/4} + 678 q^{11/4} + \mathcal{O}(q^{13/4}) \right) \,,\\
Y^{(33)}_{\mathbf{\hat{1}},3} &\simeq 4.61 \times 10^{-7} \left(q^{3/4} - 306 q^{7/4} - 320409 q^{11/4} + \mathcal{O}(q^{13/4}) \right) \,,\\
Y^{(33)}_{\mathbf{\hat{1}'},1} &\simeq 18.0 \left(q^{9/4} + \mathcal{O}(q^{13/4}) \right) \,,\\
Y^{(33)}_{\mathbf{\hat{1}'},2} &\simeq 1.63 \times 10^{-3} \left(q^{5/4} + 186 q^{9/4} + \mathcal{O}(q^{13/4}) \right) \,,\\
Y^{(33)}_{\mathbf{\hat{1}'},3} &\simeq 1.07 \times 10^{-14} \left(q^{1/4} + 930 q^{5/4} + 128961 q^{9/4} + \mathcal{O}(q^{13/4}) \right) \,,\\
Y^{(34)}_{\mathbf{1'},1} &\simeq 92.1 \left(q^{5/2} + \mathcal{O}(q^{7/2}) \right) \,,\\
Y^{(34)}_{\mathbf{1'},2} &\simeq 0.0162 \left(q^{3/2} - 804 q^{5/2} + \mathcal{O}(q^{7/2}) \right) \,,\\
Y^{(34)}_{\mathbf{1'},3} &\simeq 2.19 \times 10^{-10} \left(\sqrt{q} - 60 q^{3/2} - 392058 q^{5/2} + \mathcal{O}(q^{7/2}) \right) \,,\\
Y^{(35)}_{\mathbf{\hat{1}},1} &\simeq 0.120 \left(q^{7/4} - 66 q^{11/4} + \mathcal{O}(q^{13/4}) \right) \,,\\
Y^{(35)}_{\mathbf{\hat{1}},2} &\simeq 6.68 \times 10^{-8} \left(q^{3/4} + 678 q^{7/4} - 47025 q^{11/4} + \mathcal{O}(q^{13/4}) \right) \,,\\
Y^{(35)}_{\mathbf{\hat{1}'},1} &\simeq 8.85 \left(q^{9/4} + \mathcal{O}(q^{13/4}) \right) \,,\\
Y^{(35)}_{\mathbf{\hat{1}'},2} &\simeq 3.94 \times 10^{-4} \left(q^{5/4} + 1170 q^{9/4} + \mathcal{O}(q^{13/4}) \right) \,,\\
Y^{(35)}_{\mathbf{\hat{1}'},3} &\simeq 5.17 \times 10^{-16} \left(q^{1/4} + 186 q^{5/4} - 403191 q^{9/4} + \mathcal{O}(q^{13/4}) \right) \,,\\
Y^{(36)}_{\mathbf{1'},1} &\simeq 31.6 \left(q^{5/2} + \mathcal{O}(q^{7/2}) \right) \,,\\
Y^{(36)}_{\mathbf{1'},2} &\simeq 4.61 \times 10^{-3} \left(q^{3/2} + 180 q^{5/2} + \mathcal{O}(q^{7/2}) \right) \,,\\
Y^{(36)}_{\mathbf{1'},3} &\simeq 2.06 \times 10^{-11} \left(\sqrt{q} + 924 q^{3/2} + 123390 q^{5/2} + \mathcal{O}(q^{7/2}) \right) \,.
\end{align*}
\endgroup
%

\vskip -1cm
${ }$


\acrodef{SM}[SM]{Standard Model}
\acrodef{VEV}[VEV]{vacuum expectation value}
\acrodef{gCP}[gCP]{generalized CP}
\acrodef{CPV}[CPV]{CP violation}
\acrodef{CKM}[CKM]{Cabibbo--Kobayashi--Maskawa}
\acrodef{LR}[LR]{left-right}

\def\bibsection{\section*{\refname}}
\bibliography{bibliography}

\end{document}